\documentclass[trackchanges]{aastex701}

\newcommand\tess{\textit{TESS}}
\newcommand\kepler{\textit{Kepler}}

\newcommand{\Rbr}{\ensuremath{9.2_{-1.1}^{+1.9}}}

\newcommand{\Lam}{\ensuremath{16.4_{-9.7}^{+8.1}}}

\newcommand{\muOne}{\ensuremath{0.039_{-0.038}^{+0.018}}}

\newcommand{\muTwo}{\ensuremath{0.466_{-0.068}^{+0.067}}}

\newcommand{\kapOne}{\ensuremath{18_{-17}^{+11}}}

\newcommand{\kapTwo}{\ensuremath{8.5_{-4.9}^{+3.3}}}

\newcommand{\wmin}{\ensuremath{0.22_{-0.15}^{+0.12}}}

\newcommand{\wmax}{\ensuremath{0.810_{-0.069}^{+0.078}}}

\newcommand{\wLowSN}{\ensuremath{0.808_{-0.067}^{+0.079}}}
\newcommand{\wHighSN}{\ensuremath{0.192_{-0.079}^{+0.067}}}
\newcommand{\muMixSN}{\ensuremath{0.127_{-0.033}^{+0.033}}}
\newcommand{\wLowSS}{\ensuremath{0.78_{-0.08}^{+0.11}}}
\newcommand{\wHighSS}{\ensuremath{0.22_{-0.11}^{+0.08}}}
\newcommand{\muMixSS}{\ensuremath{0.142_{-0.038}^{+0.032}}}
\newcommand{\wLowJOV}{\ensuremath{0.35_{-0.13}^{+0.12}}}
\newcommand{\wHighJOV}{\ensuremath{0.65_{-0.12}^{+0.13}}}
\newcommand{\muMixJOV}{\ensuremath{0.316_{-0.042}^{+0.041}}}

\usepackage{amsmath} 
\usepackage{multirow}

\begin{document}

\title{The Orbital Eccentricity--Radius Distribution for Warm, Single Planets in \tess}

\author[orcid=0000-0002-0692-7822,sname='Fairnington']{Tyler R. Fairnington}
\affiliation{Centre for Astrophysics, University of Southern Queensland, Toowoomba, QLD 4350, Australia}
\affiliation{Department of Astronomy \& Astrophysics, University of Chicago, Chicago, IL 60637, USA}
\email[show]{tfairnington@uchicago.edu}

\author[orcid=0000-0002-3610-6953,gname=Jiayin,sname=Dong]{Jiayin Dong}
\affiliation{Department of Astronomy, University of Illinois at Urbana-Champaign, Urbana, IL 61801, USA}
\affiliation{Center for Astrophysical Surveys, National Center for Supercomputing Applications, Urbana, IL 61801, USA}
\email[]{jdongx@illinois.edu}

\author[orcid=0000-0003-0918-7484,sname='Huang']{Chelsea X. Huang}
\affiliation{Centre for Astrophysics, University of Southern Queensland, Toowoomba, QLD 4350, Australia}
\email[]{chelsea.huang@unisq.edu.au}

\author[orcid=0000-0003-0571-2245,sname='Nabbie']{Emma Nabbie}
\affiliation{Centre for Astrophysics, University of Southern Queensland, Toowoomba, QLD 4350, Australia}
\email[]{emma.nabbie@unisq.edu.au}

\author[orcid=0000-0002-4891-3517,sname='Zhou']{George Zhou}
\affiliation{Centre for Astrophysics, University of Southern Queensland, Toowoomba, QLD 4350, Australia}
\email[]{george.zhou@unisq.edu.au}

\author[orcid=0000-0001-7294-5386,sname='Wright']{Duncan Wright}
\affiliation{Centre for Astrophysics, University of Southern Queensland, Toowoomba, QLD 4350, Australia}
\email[]{duncan.wright@unisq.edu.au}

\author[orcid=0000-0001-6588-9574]{Karen A.\ Collins}
\affiliation{Center for Astrophysics \textbar \ Harvard \& Smithsonian, 60 Garden Street, Cambridge, MA 02138, USA}
\email[]{karen.collins@cfa.harvard.edu}

\author[orcid=0000-0002-5741-3047]{David Ciardi}
\affiliation{NASA Exoplanet Science Institute-Caltech/IPAC, Pasadena, CA 91125, USA}
\email[]{ciardi@ipac.caltech.edu}

\author[orcid=0000-0002-4715-9460,sname='Jenkins']{Jon M. Jenkins}
\affiliation{NASA Ames Research Center, Moffett Field, CA 94035, USA}
\email[]{jon.jenkins@nasa.gov}

\author[orcid=0000-0001-9911-7388,sname='Latham']{David W. Latham}
\affiliation{Center for Astrophysics \textbar \ Harvard \& Smithsonian, 60 Garden Street, Cambridge, MA 02138, USA}
\email[]{dlatham@cfa.harvard.edu}

\author[orcid=0000-0003-2058-6662,sname='Ricker']{George Ricker}
\affiliation{Department of Physics and Kavli Institute for Astrophysics and Space Research, Massachusetts Institute of Technology, 77 Massachusetts Ave, Cambridge, MA 02139, USA}
\email[]{grr@space.mit.edu}

\author[orcid=0000-0002-8964-8377,sname='Quinn']{Samuel N. Quinn}
\affiliation{Center for Astrophysics \textbar \ Harvard \& Smithsonian, 60 Garden Street, Cambridge, MA 02138, USA}
\email[]{squinn@cfa.harvard.edu}

\author[orcid=0000-0002-6892-6948,sname='Seager']{Sara Seager}
\affiliation{Department of Earth, Atmospheric and Planetary Sciences, Massachusetts Institute of Technology, 77 Massachusetts Avenue, Cambridge, MA 02139, USA}
\affiliation{Department of Physics and Kavli Institute for Astrophysics and Space Research, Massachusetts Institute of Technology, 77 Massachusetts Ave, Cambridge, MA 02139, USA}
\affiliation{Department of Aeronautics and Astronautics, Massachusetts Institute of Technology, Cambridge, MA 02139, USA}
\email[]{profseager@mit.edu}

\author[orcid=0000-0002-1836-3120, sname='Shporer']{Avi Shporer}
\affiliation{Department of Physics and Kavli Institute for Astrophysics and Space Research, Massachusetts Institute of Technology, 77 Massachusetts Ave, Cambridge, MA 02139, USA}
\email[]{shporer@mit.edu}

\author[orcid=0000-0001-6763-6562,sname='Vanderspek']{Roland Vanderspek}
\affiliation{Department of Physics and Kavli Institute for Astrophysics and Space Research, Massachusetts Institute of Technology, 77 Massachusetts Ave, Cambridge, MA 02139, USA}
\email[]{roland@space.mit.edu}

\author[orcid=0000-0002-4265-047X,sname='Winn']{Joshua N. Winn}
\affiliation{Department of Astrophysical Sciences, Princeton University, Princeton, NJ 08544, USA}
\email[]{jnwinn@princeton.edu}

\author[orcid=0000-0003-1464-9276]{Khalid Barkaoui}
\affiliation{Instituto de Astrofísica de Canarias (IAC), E-38200 La Laguna, Tenerife, Spain}
\affiliation{Astrobiology Research Unit, Universit\'e de Li\`ege, All\'ee du 6 Ao\^ut 19C, B-4000 Li\`ege, Belgium}
\affiliation{Department of Earth, Atmospheric and Planetary Sciences, Massachusetts Institute of Technology, 77 Massachusetts Avenue, Cambridge, MA 02139, USA}
\email{khalid.barkaoui@uliege.be}

\author[orcid=0000-0001-6637-5401]{Allyson Bieryla}
\affiliation{Center for Astrophysics \textbar \ Harvard \& Smithsonian, 60 Garden Street, Cambridge, MA 02138, USA}
\email[]{abieryla@cfa.harvard.edu}

\author[orcid=0000-0003-1605-5666]{Lars Buchhave}
\affiliation{DTU Space, Technical University of Denmark, Elektrovej 328, DK-2800 Kgs. Lyngby, Denmark}
\email[]{buchhave@space.dtu.dk}

\author[orcid=0009-0003-4203-9667]{Dmitry Cheryasov}
\affiliation{Sternberg Astronomical Institute, Lomonosov Moscow State University, Moscow, Russia, 119992}
\email[]{dmitry.cheryasov@gmail.com}

\author[orcid=0000-0002-8035-4778]{Jessie Christiansen}
\affiliation{NASA Exoplanet Science Institute-Caltech/IPAC, Pasadena, CA 91125, USA}
\email[]{jessie.christiansen@caltech.edu}

\author[orcid=0000-0001-8189-0233]{Courtney Dressing}
\affiliation{Department of Astronomy, 501 Campbell Hall \#3411, University of California, Berkeley, CA 94720, USA}
\email[]{dressing@berkeley.edu}

\author[orcid=0000-0002-4909-5763]{Akihiko Fukui}
\affiliation{Komaba Institute for Science, The University of Tokyo, 3-8-1 Komaba, Meguro, Tokyo 153-8902, Japan}
\affiliation{Instituto de Astrof\'{i}sica de Canarias (IAC), 38205 La Laguna, Tenerife, Spain}
\email[]{afukui@g.ecc.u-tokyo.ac.jp}

\author[orcid=0000-0003-2599-1405]{Alexey Garmash}
\affiliation{Novosibirsk State University, Novosibirsk 630090, Russia}
\email[]{garmash.alexey@gmail.com}

\author[orcid=0000-0002-8965-3969]{Steven Giacalone}
\affiliation{Department of Astronomy, California Institute of Technology, Pasadena, CA 91125, USA}
\email[]{steven_giacalone@berkeley.edu}

\author[orcid=0000-0002-9867-7938]{Eric G. Hintz}
\affiliation{Department of Physics and Astronomy, Brigham Young University, N-486 ESC, Provo, UT 84602 USA}
\email[]{hintz@byu.edu}

\author[orcid=0000-0002-2532-2853]{Steve B. Howell}
\affiliation{NASA Ames Research Center, Moffett Field, CA 94035, USA}
\email[]{steve.b.howell@nasa.gov}

\author[orcid=0000-0002-6480-3799]{Keisuke Isogai}
\affiliation{Okayama Observatory, Kyoto University, 3037-5 Honjo, Kamogatacho, Asakuchi, Okayama 719-0232, Japan}
\affiliation{Department of Multi-Disciplinary Sciences, Graduate School of Arts and Sciences, The University of Tokyo, 3-8-1 Komaba, Meguro, Tokyo 153-8902, Japan}
\email[]{isogai@kusastro.kyoto-u.ac.jp}

\author[orcid=0000-0002-6424-3410]{Jerome de Leon}
\affiliation{Komaba Institute for Science, The University of Tokyo, 3-8-1 Komaba, Meguro, Tokyo 153-8902, Japan}
\email[]{jpdeleon@g.ecc.u-tokyo.ac.jp}

\author[orcid=0000-0003-3742-1987]{Jorge Lillo-Box}
\affiliation{Centro de Astrobiología, CSIC-INTA, Camino Bajo del Castillo s/n, 28692 Villanueva de la Cañada, Madrid, Spain}
\email[]{jlillo@cab.inta-csic.es}

\author[orcid=0000-0001-9087-1245]{Felipe Murgas}
\affiliation{Instituto de Astrof\'isica de Canarias (IAC), E-38205 La Laguna, Tenerife, Spain}
\affiliation{Departamento de Astrof\'isica, Universidad de La Laguna (ULL), E-38206 La Laguna, Tenerife, Spain}
\email[]{fmurgas@iac.es}

\author[orcid=0000-0001-8511-2981]{Norio Narita}
\affiliation{Komaba Institute for Science, The University of Tokyo, 3-8-1 Komaba, Meguro, Tokyo 153-8902, Japan}
\affiliation{Astrobiology Center, 2-21-1 Osawa, Mitaka, Tokyo 181-8588, Japan}
\affiliation{Instituto de Astrof\'{i}sica de Canarias (IAC), 38205 La Laguna, Tenerife, Spain}
\email[]{narita@g.ecc.u-tokyo.ac.jp}

\author[orcid=0000-0002-5254-2499]{Louise D. Nielsen}
\affiliation{Observatoire de Genève, Département d’Astronomie, Université de Genève, Chemin Pegasi 51b, 1290 Versoix, Switzerland}
\affiliation{University Observatory, Faculty of Physics, Ludwig-Maximilians-Universität München, Scheinerstr. 1, 81679 Munich, Germany}
\email[]{Louise.Nielsen@lmu.de}

\author[orcid=0000-0003-0987-1593]{Enric Palle}
\affiliation{Instituto de Astrof\'{i}sica de Canarias (IAC), 38205 La Laguna, Tenerife, Spain}
\affiliation{Departamento de Astrof\'isica, Universidad de La Laguna (ULL), E-38206 La Laguna, Tenerife, Spain}
\email[]{epalle@iac.es}

\author[orcid=0000-0003-2935-7196]{Markus Rabus}
\affiliation{Departamento de Matemática y Física Aplicadas, Facultad de Ingeniería, Universidad Católica de la Santísima Concepción, Alonso de Rivera 2850, Concepción, Chile}
\email[]{mrabus@lco.global}

\author[orcid=0000-0002-3627-1676]{Benjamin V.\ Rackham}
\affiliation{Department of Earth, Atmospheric and Planetary Sciences, Massachusetts Institute of Technology, 77 Massachusetts Avenue, Cambridge, MA 02139, USA}
\affiliation{Kavli Institute for Astrophysics and Space Research, Massachusetts Institute of Technology, Cambridge, MA 02139, USA}
\email[]{brackham@mit.edu}

\author[orcid=0000-0001-8227-1020]{Richard P. Schwarz}
\affiliation{Center for Astrophysics \textbar \ Harvard \& Smithsonian, 60 Garden Street, Cambridge, MA 02138, USA}
\email[]{rpschwarz@comcast.net}

\author[]{Gregor Srdoc}
\affiliation{Kotizarovci Observatory, Sarsoni 90, 51216 Viskovo, Croatia}
\email[]{gregorsrdoc@gmail.com}

\author[orcid=0000-0003-4658-7567]{Denise C. Stephens}
\affiliation{Department of Physics and Astronomy, Brigham Young University, N-486 ESC, Provo, UT 84602 USA}
\email[]{denise_stephens@byu.edu}

\author[orcid=0000-0003-3092-4418]{Gavin Wang}
\affiliation{William H. Miller III Department of Physics \& Astronomy, Johns Hopkins University, Baltimore, MD 21218, USA}
\email[]{gxwang22@gmail.com}

\author[orcid=0000-0002-7522-8195]{Noriharu Watanabe}
\affiliation{Department of Multi-Disciplinary Sciences, Graduate School of Arts and Sciences, The University of Tokyo, 3-8-1 Komaba, Meguro, Tokyo 153-8902, Japan}
\email[]{n-watanabe@g.ecc.u-tokyo.ac.jp}

\author[orcid=0000-0003-2127-8952]{Francis P. Wilkin}
\affiliation{Department of Physics and Astronomy, Union College, 807 Union St., Schenectady, NY 12308, USA}
\email[]{wilkinf@union.edu}

\author[orcid=0009-0006-0339-7551]{Joe Williams}
\affiliation{Department of Physics and Astronomy, Brigham Young University, N-486 ESC, Provo, UT 84602 USA}
\email[]{jbwill2019@outlook.com}

\begin{abstract}

We characterize the radius-dependent observed transiting eccentricity distribution of 219 warm ($P = 8$--50~days) systems with only one transiting planetary candidate identified during Sectors 1--69 of the \tess{} mission. Using the ``photoeccentric effect'' in a hierarchical Bayesian framework, we first model the population using discrete planetary size bins (sub-Neptunes, sub-Saturns, and Jovians). We then develop a continuous mixture model with weights governed by a logistic sigmoid function of radius. We find that the warm-single population is best described by two components: a dominant low-eccentricity mode ($\langle e_{\rm low}\rangle=$ \muOne) and a secondary dynamically excited mode ($\langle e_{\rm high}\rangle=$ \muTwo). The fraction of planets belonging to this high-eccentricity component increases strongly with planet radius, characterized by a transition at a break radius of $R_{\rm br}=$ \Rbr{} $R_\oplus$. This trend places warm sub-Saturns predominantly on the same low-eccentricity track as sub-Neptunes.
In contrast, warm Jovians (8--16$\,R_\oplus$) are frequently eccentric, with $65_{-12}^{+13}\%$ of the population in the high eccentricity mode.  Under the assumption of a two-component model, we see tentative evidence for a bimodal Jovian distribution at ${\sim}$2.7$\sigma$.
Finally, we identify a non-negligible tail of highly eccentric sub-Neptunes (1--4$\,R_\oplus$), which comprise $16.2_{-6.4}^{+5.2}\%$ of the population, consistent with excitation by non-transiting external companions.
\end{abstract}


\section{Introduction} 

Orbital eccentricity encodes the formation and dynamical evolution of planetary systems. 
Disk migration and \textit{in-situ} formation are expected to yield nearly circular orbits at the birth of planets. In contrast, post-disk dynamical processes---including planet--planet scattering and secular excitation (e.g., von Zeipel--Kozai--Lidov cycles and secular chaos)---can generate substantial eccentricities (e.g., \citealt{Rasio&Ford:1996, Chatterjee:2008, Kozal:1962, Lidov:1962, Naoz:2016}; see \citealt{Dawson:2018} for a review).
Observations of the eccentricity distribution are especially valuable for ``warm'' planets ($P\simeq8-50$~days) as their dynamical imprints are often preserved. Tidal circularization is typically inefficient for warm planets and their eccentricities more closely reflect the primordial values and the outcome of post-formation evolution \citep{Eggleton:1998, Dawson:2018}. Characterizing the eccentricity distribution of warm planets therefore provides direct constraints on the relative importance of quiescent versus dynamically excited evolutionary pathways. 

Transit surveys enable population-scale eccentricity studies across a wide range of planet sizes using the ``photoeccentric effect''. The technique leverages the dependence of the light curve duration and shape on stellar density and orbital velocity, as a planet transiting near periastron (apoastron) moves faster (slower) than a circular analog \citep{Dawson:2012, Seager:2003}. Early analyses of \kepler{} samples using this technique established that single-transiting (``singles'') systems are typically significantly more eccentric than compact multi-planet (``multis'') systems \citep{Xie:2016, VanEylen:2015, VanEylen:2019}, consistent with singles preferentially tracing dynamically active architectures. 

However, individual eccentricity measurements are often highly uncertain due to degeneracies with the argument of periastron and uncertainties in stellar parameters. To robustly recover population distributions from these noisy individual measurements, recent work has increasingly adopted Hierarchical Bayesian Modeling \citep[HBM;][]{Hogg:2010, DFM:2014}. Using this framework, \citet{VanEylen:2019} refined the eccentricity distributions of \kepler{} small planets, while \citet{Gilbert:2025} conducted the first radius--eccentricity relation in a hierarchical model. More recently, analyses with NASA's \textit{Transiting Exoplanet Survey Satellite} (\tess) mission \citep{Ricker:2015} of warm Jovians and sub-Saturns have revealed evidence for bimodal eccentricity distributions, suggesting diverse formation and dynamical evolution pathways \citep{Dong:2021,Fairnington:2025}. 

Despite these advances, \tess{} results have not been unified into a single homogeneous analysis spanning the full planet-radius range. Existing studies have typically been segmented by planet size categories---focusing on Jovians or sub-Saturns. This makes it difficult to trace how dynamical excitation evolves continuously from the sub-Neptunes to Jovians in the same observed population. Furthermore, while \kepler{} studies have explored radius trends (e.g., \citealt{Gilbert:2025}), the \tess{} yield of warm planets benefits from all-sky coverage, offering a statistically robust sample of these intrinsically rare objects that is distinct from the \kepler{} field and more amenable to future follow-up. 

In this work, we present the first \tess-based population-level inference of the eccentricity distribution for warm, single-transiting (``single'') planets across the full radius range. By ``single'' we mean that only one transiting planet has been detected around the star. We construct a homogeneous sample of 219 warm singles from the \tess{} Prime and Extended Missions, derive precise stellar densities from Gaia DR3 \citep{gaiadr2, GaiaDR3} and independent stellar modeling, and infer the eccentricity distribution using HBM. To connect sub-Neptunes through gas giants within a single framework, we introduce a radius-continuous three-stage hierarchical mixture model that propagates radius uncertainties and quantifies how the eccentricity distribution evolves with planet size. The paper is organized as follows: Section \S\ref{sec:methods} describes the sample compilation and light curve processing; Section \S\ref{sec:analysis} details the fitting of stellar and planetary parameters; Section \S\ref{sec:results} presents the inferred eccentricity distributions and their radius dependence; and Section \S\ref{sec:discussion} interprets these results in the context of the literature. 

\section{Data}

\label{sec:methods}
\subsection{Initial Sample}

We analyze the 6881 readily available planet candidates from the \tess{} Object of Interest (TOI, \citealt{TOI}) catalog as at the conclusion of the second extended mission\footnote{Retrieved from the Exoplanet Archive on 2025-01-05 UT \citep{NEA}}. This corresponds to all candidates identified through Sectors 1--69. We remove all candidates with host star brightness fainter than \tess-band magnitudes of 12. We restrict our sample to bright stars because high photometric precision is required to resolve the transit ingress and egress durations, which are critical for breaking the degeneracy between the impact parameter and the stellar density. We discuss the effect of SNR more in Section \S\ref{subsection{SNR}}. We also remove targets with a Gaia Renormalized Unit Weight Error (RUWE) of $\ge1.4$ \citep{Wood:2021, gaiadr2, GaiaDR3}, as they are more likely to be associated with stellar binarity. We aim to characterize targets around main-sequence stars, and thus remove hosts with log(g)$<$4. Due to the influence of planet-star interactions at short orbital periods, we restrict our sample to ``warm" planets--those with orbital periods between 8 and 50 days--where the lower limit is the approximate period where tidal circularization falls off \citep{Jackson:2008, Millholland:2025}, and the upper serves as an estimation to the window function corresponding to two sectors of \tess{} coverage. We also remove targets with False Positive (FP), False Alarm (FA) and/or Ambiguous Planet Candidate (APC) dispositions as reported by the \tess{} vetting team and/or \tess{} Follow-Up Observing Program (TFOP). As our primary motivation for this study is to reveal the dynamical origins within the exoplanet population, in paper I, we focus on single planets. Our initial sample therefore consists of 539 targets, 110 of which are confirmed and/or validated planets ($\approx$20\% of our sample)\footnote{HIP 113103c (TIC 121490076) was removed as it is a confirmed multi-transiting system \citep{Lowson:2024} labeled as a single-TOI system}. 

\subsection{Light Curve Retrieval}
Individual light curves are retrieved and processed from Mikulski Archive for Space Telescopes (MAST) with the \texttt{Lightkurve} package \citep{lightkurve}. Each target uses data up to Sector 69 of \tess, where applicable. We take the shortest cadence for each sector, corresponding to 2-min cadence where the Science Processing Operations Center (SPOC, \citealt{SPOC1, SPOC2}) data is available, and 200-sec, 10-min or 30-min cadence for data processed by the MIT Quick-Look Pipeline (QLP, \citealt{QLP}). In all instances, we use the Simple Aperture Photometry (SAP) flux. We correct for contamination in SPOC SAP data (QLP SAP is already decontaminated). We enforce all quality flags to exclude any cadences affected by systematics and/or scattered light. We derive our own flux error for every dataset using 1.4826 $\times$ the median absolute deviation (MAD) of the flux for each individual sector, as a robust method of characterizing noise in datasets with outliers (this method is standard in QLP). We do not perform detrending on the raw light curves at this stage. Light curves are simultaneously detrended with global fitting, as described in Section \S\ref{subsec:planet}. 

\subsection{Signal-to-Noise Cuts}

\label{subsection{SNR}}

\begin{figure}
\centering{}
    \includegraphics[width=0.49\textwidth]{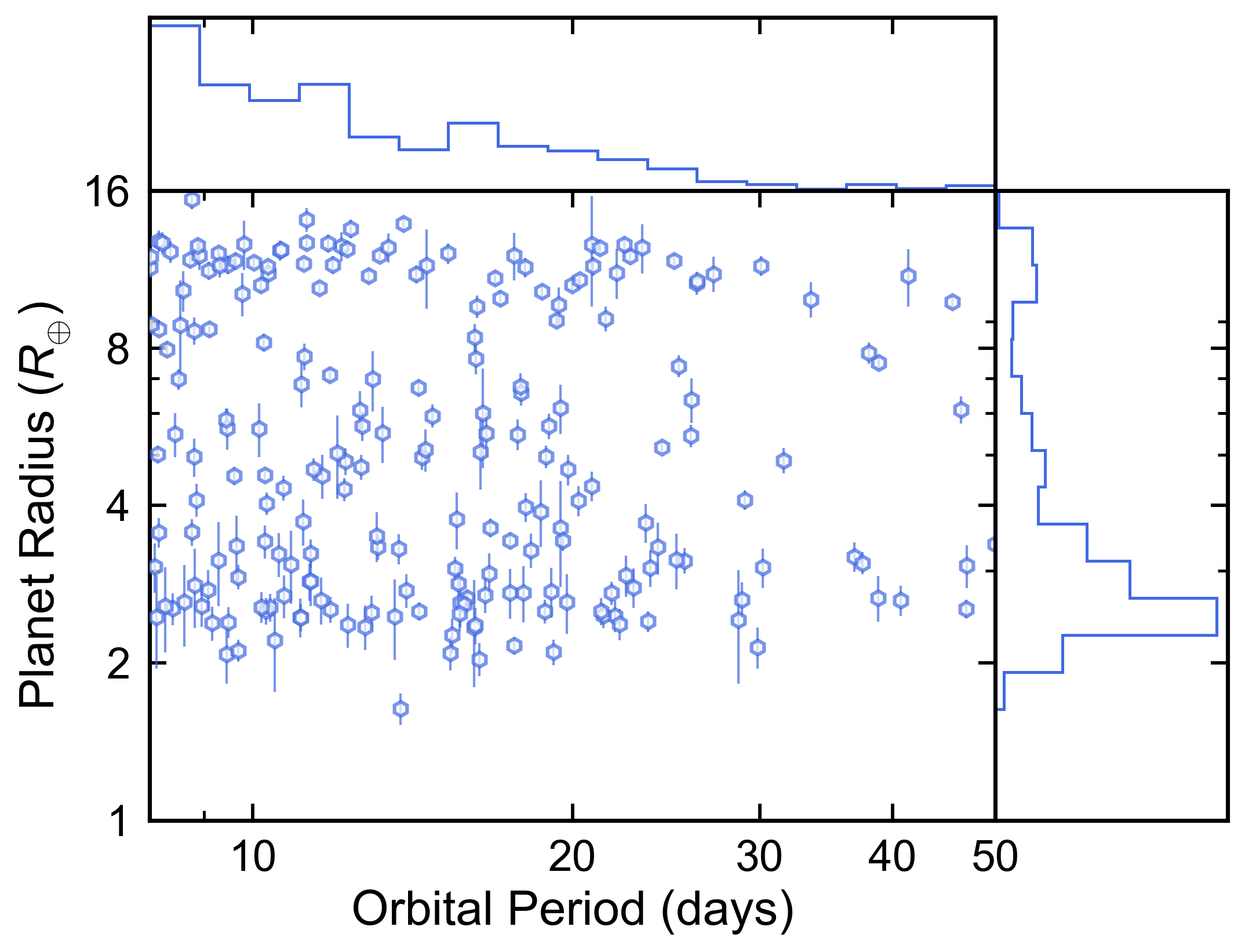}
    \caption{Planet radius versus orbital period for the final sample. Points and error bars represent median posterior values and 68\% credible intervals. The side panels show the 1-D histograms for radius (right) and period (top).}
    \label{fig:periodradius}
\end{figure}

We require sufficient Signal-to-Noise (SNR) targets to derive a reliable eccentricity, and to identify Transit Timing Variations (TTVs), which may otherwise compromise the derived eccentricity \citep{Kipping:2013, Kipping:2025}. We therefore restrict our study to targets with SNR's above 15. As fitting all planets and deriving their SNR's post-analysis is computationally inefficient, we apply a conservative approximation cut by requiring planets to satisfy one of two noise criteria. The first is a theoretical noise floor, derived from a characteristic polynomial of the star's magnitude ($T_{\mathrm{mag}}$) that accounts for photon limits and instrumental read noise of \tess{} \citep{Cooke:2018}. Alternatively, we calculate an empirical conservative noise metric by computing the standard deviation of the shortest-cadence data in a given sector and scaling the photometric noise to a one-hour noise estimate. We then apply the standard SNR formula using the planet-to-star radius ratio and transit duration from the TOI catalog, as well as the expected number of transits in the dataset by propagating the TOI period and epoch through a linear ephemeris through all observed sectors. Candidates are removed if they fail both of these estimates. This resulted in the removal of 182 low-SNR candidates. We perform additional reductions of the sample in Section \S\ref{subsec:planet} to exclude targets where their orbital characteristics are unconstrained or problematic for deriving eccentricity (i.e. period aliases and/or grazing transits). The final sample can be seen in Figure \ref{fig:periodradius}.

\section{Analysis}

\label{sec:analysis}

Our analysis proceeds in three stages. First, we derive independent constraints on the true stellar density, $\rho_\star$, using spectral energy distribution (SED) and isochrone fitting (or empirical relations for cool stars). Second, we fit each \tess{} transit light curve to measure the ``pseudo'' stellar density inferred under a circular-orbit assumption, $\tilde{\rho}$, alongside the standard transit parameters. Third, we combine $\tilde{\rho}$ and $\rho_\star$ to infer a joint $(e,\omega)$ posterior for each planet via the photoeccentric effect, prior to our hierarchical modeling.

\subsection{Stellar density inference from SED and isochrone fitting}
\label{subsec:stellar}

Independent constraints on $\rho_\star$ are required to derive eccentricity from a transit light curve via the photoeccentric effect \citep{FQV:08, Dawson:2012, Kipping:2010}. We therefore perform a uniform stellar analysis for the full sample using \texttt{astroARIADNE} to model each star's SED, and then map the resulting stellar parameters onto \textit{Mesa Isochrones and Stellar Tracks} (MIST) isochrones to infer $\rho_\star$.

We include a variety of the available stellar atmospheric models (Phoenixv2, \citealt{Husser:2013}; BT-Settl, \citealt{Allard:2012}; BT-NextGen, \citealt{Allard:2012}; BT-Cond, \citealt{Hauschildt:1999, Allard:2012}; Kurucz93, \citealt{Kurucz:1993}) in our fits, leveraging Bayesian Model Averaging to weight the best-fitting posteriors. These results are then fed into the MIST isochrones to infer the best-fitting stellar density. Each SED is composed of photometry from the Gaia DR3 $G$, $Bp$ and $Rp$ mags \citep{GaiaDR3}, Two-Micron All-Sky Survey (2MASS) $J$, $H$ and $K$ \citep{2MASS}, Wide field Infrared Survey Explorer (WISE) $W_1$ and $W_2$ \citep{WISE}, as well as Tycho-2 B and V where available \citep{Tycho2}, resorting to the Johnson magnitudes\footnote{If the Johnson V magnitude is within 3 magnitudes of Gaia G magnitude} where Tycho-2 is unavailable. We incorporate Gaia DR3 parallax measurements as a prior on the distance, supplemented with a prior on extinction using the galactic dust maps by \citet{Avdustmap}. Other stellar parameters, namely [Fe/H], $\mathrm{T_{eff}}$ and log(g), all have the \texttt{astroARIADNE} default empirical priors from the Radial Velocity (RAVE) survey \citep{RAVE}.

SED and subsequent isochrone inference are known to be less reliable for cool stars. For targets with $T_{\mathrm{eff}} \le 4000$~K, we instead adopt empirical relations that yield more precise densities for M dwarfs. We infer stellar masses using the $M_\star$--$K_s$ relation from \citet{Mann:2019}, and stellar radii using the probabilistic mass--radius relation from \citet{Kipping:2025}, from which we compute $\rho_\star$. To prevent biases induced by imprecise stellar parameters, we perform two conservative cuts, namely removing all targets with stellar effective temperature uncertainties exceeding 250 K, which removed 19 targets; and removal of targets with host star density uncertainties exceeding 20\%, which removed 10 targets. 

\subsection{Transit light-curve modeling and pseudo-density constraints}
\label{subsec:planet}

We constrain the eccentricity of each planet via the ``photoeccentric'' effect \citep{Dawson:2012}. This technique compares the host star's true density to the density derived from the light curve under the assumption of a circular orbit \citep{Seager:2003}. Since the light-curve density is determined by the transit duration and impact parameter, it will deviate from the true density if the orbit is eccentric. 
We refer to this as the ``pseudo'' stellar density, $\tilde{\rho}$ attributed to the assumptions extending beyond simply following a Keplerian orbit \citep{Kipping:2013, Gilbert:2022}. Given \textit{a priori} knowledge of the ``true'' stellar density, $\rho_{\star}$, which is enabled through Section \S\ref{subsec:stellar}, the eccentricity of a planet can be derived as 

\begin{align}
\label{eqn:photoeccentric}
    \mathrm{g}(e,\omega)^3 = \frac{\tilde{\rho}(e,\omega)}{\rho_{\star}(e,\omega)}
\end{align}
where 
\begin{align}
    \mathrm{g}(e, \omega) = \frac{1 + e\ \mathrm{sin}\omega}{\sqrt{1-e^2}}
\end{align}
represents the ratio of the planets velocity during transit to that of its expected circular velocity for a given orbital period \citep{Dawson:2012, Kipping:2010}. We thus compare the true stellar density, obtained from Section \S\ref{subsec:stellar}, to that of the pseudo stellar density by jointly modeling the eccentricity and argument of periastron through the photoeccentric relation, as described in Section \S\ref{subsec:individualeccentricities}.

To obtain precise constraints of $\tilde\rho$, as well as the individual planets' orbital parameters, we perform light curve modeling. The procedure largely follows that of \citet{Fairnington:2025}. In brief: we simultaneously fit the planet transits with a third-order polynomial function of time to account for instrumental and astrophysical noise. To mitigate biases introduced by finite integration time and heterogeneous observing cadence, all light curves are supersampled during model evaluation following the prescription of \citet{Kipping:2010}. To increase computational efficiency, we reduce the light curves to include only data within three transit durations of the expected transit time.  For systems with potential TTVs (identified from visual inspection of the raw light curve and/or from residual transit smearing after an initial linear-ephemeris fit), we first manually assign all mid-transit points before applying the truncation. The free parameters in our model consist of the planet's orbital period, $P$, reference transit center, $T_c$, impact parameter, $b$, the planet-to-star radius ratio, $R_p/R_\star$, the pseudo stellar density, $\tilde{\rho_\star}$, and the \citet{Kipping:LD} reparameterization of the limb-darkening coefficients, $q_1,q_2$. We adopt uniform priors on the limb-darkening parameters, following \citet{Kipping:LD}, so that limb darkening is fit freely over the physically allowed quadratic-law parameter space. We prefer this conservative choice over stellar-model-informed priors in order to avoid introducing additional model dependence and/or biases to the sample \citep{EspinozaJordan:2015}.
For the default fit, we assume a linear ephemeris and fit only $P$ and $T_c$, rather than allowing all individual transit times to vary independently. We apply this prescription to all systems as an initial pass, and then manually inspect the individual transits and phase-folded light curves. For systems that show clear evidence of TTVs or transit smearing under the best-fit linear ephemeris, we instead allow the individual transit times, $TT_i$, to float and derive a linear approximation post hoc from a least-squares fit to the recovered transit times. Because fitting free $TT_i$ for every target in the sample is computationally prohibitive, we reserve this treatment for systems that fail the linear-ephemeris vetting. We note that either choice may introduce bias if low-amplitude TTVs remain undetected, and therefore treat this as a caveat of the analysis.
We fit the pseudo stellar density in log-space to prevent biases between the b-$\tilde{\rho_\star}$ covariance \citep{Gilbert:2022}. The full list of priors can be seen in Table \ref{tab:priors}. 

We perform modeling with Hamiltonian Monte Carlo (HMC) using the \texttt{PyMC} package \citep{pymc5}, constructing a quadratically limb-darkened light curve with the \texttt{exoplanet} package. We use the No-U-Turn Sampler \citep{NUTS} with 5000 tuning steps and a subsequent 5000 draws with a target acceptance of 0.99 to limit degeneracies from modeling the pseudo stellar density \citep{Dong:2021}. We fit four chains simultaneously, with convergence satisfied through a Gelman-Rubin diagnostic \citep{Rhat} $\mathcal{\hat{R}}\le1.01$, as well as manual vetting of the corner plots, posterior chains and phase-folded light curve plots. 

We remove targets with unresolved period aliases, as these compromise the interpretation of the planet's eccentricity (17 targets removed), those with $>50\%$ of samples with $b\ge0.9$, due to their highly degenerate posterior space and unconstrained radii (48 targets removed), and planets with radius uncertainties exceeding 25\% (17 removed). 219 planets are therefore used in our final eccentricity distribution. 

\begin{deluxetable}{lcc}
\tablecaption{Priors adopted for the transit light-curve fits.\label{tab:priors}}
\tablewidth{0pt}
\tablehead{
\colhead{Parameter} & \colhead{Linear Prior} & \colhead{TTV Prior}
}
\startdata
$b$ & $\mathrm{abs}(\mathcal{U}(-2,2))$ & $\mathrm{abs}(\mathcal{U}(-2,2))$ \\
$\mathrm{log}(R_p/R_{\star})$ & $\mathcal{U}(\mathrm{log}\sqrt{30\times10^{-6}}, \mathrm{log}\sqrt{0.5})$ & $\mathcal{U}(\mathrm{log}\sqrt{30\times10^{-6}}, \mathrm{log}\sqrt{0.5})$ \\
$\mathrm{log}(\tilde{\rho}_\star)$ & $\mathcal{U}(-3, 3)$ & $\mathcal{U}(-3, 3)$ \\
$q_1, q_2$ & $\mathcal{U}(0, 1)$ & $\mathcal{U}(0, 1)$ \\
$P$ & $\mathcal{N}(\mu_{\mathrm{TOI}}, 10\sigma_{\mathrm{TOI}})$ & --- \\
$T_c$ & $\mathcal{N}(\mu_{\mathrm{TOI}}, 10\sigma_{\mathrm{TOI}})$ & --- \\
$TT_i$ & --- & $\mathcal{N}(\mu_{\mathrm{manual}}, 0.1)$ \\
\enddata
\tablecomments{``Linear'' denotes fits with a linear ephemeris (fixed transit times), while ``TTV'' denotes fits in which individual transit times $TT_i$ are free parameters. $\mathcal{U}$ and $\mathcal{N}$ denote uniform and normal priors, respectively. For $P$ and $T_c$, $\mu_{\mathrm{TOI}}$ and $\sigma_{\mathrm{TOI}}$ are the catalog-reported mean value and $1\sigma$ uncertainty from the TOI list. For TTV fits, $\mu_{\mathrm{manual}}$ is the manually identified mid-transit time used to initialize each $TT_i$ prior (in days).}
\end{deluxetable}

\subsection{Individual eccentricity constraints from the photoeccentric effect}
\label{subsec:individualeccentricities}

Prior to our population-level analysis, we first perform an individual eccentricity derivation for all targets in our final sample. With the posterior solutions for both $\tilde\rho$ and $\rho_{\star}$, we derive eccentricities following equation \ref{eqn:photoeccentric} by directly sampling the joint posterior distribution of eccentricity and argument of periastron. This approach constrains the properties of an unobserved parameter of interest, the planet eccentricity, through a related observed distribution, the pseudo stellar density, given a functional relation.
In this case, the distributions are related via $g(e,\omega)$. Notably, we constrain the joint marginal posterior of $e-\omega$, given they are degenerate in the solution \citep{Dawson:2012}. We fit for the re-parameterized $\sqrt{e}\ \mathrm{sin}\omega$ and $\sqrt{e}\ \mathrm{cos}\omega$ in our analysis, each imposed with uniform priors of [$-1, 1$]. We use the same model strategy as our planet fitting for this analysis. We also perform a post-analysis Kolmogorov-Smirnov (KS) test to ensure that the posterior eccentricity provides a meaningful constraint rather than simply reflecting an uninformative distribution. We compare the marginalized eccentricity posterior samples (truncated to the physical domain [0,0.9]) against a uniform distribution. To avoid statistical overpowering due to large chain sizes, we subsample the posterior to N=1000 draws before calculating the KS statistic. Posteriors that are statistically indistinguishable from a uniform distribution (p$>$0.05) imply that the light curve contains insufficient information to constrain the eccentricity through the transit shape. To avoid biasing the population inference, we assign these unconstrained systems a broad unconstrained value ($e=0.5\pm0.5$). We tested the removal of these 13 targets ($<4\%$ of the sample) and the results/conclusions remained unchanged. 

\begin{figure}
    \includegraphics[width=0.99\textwidth]{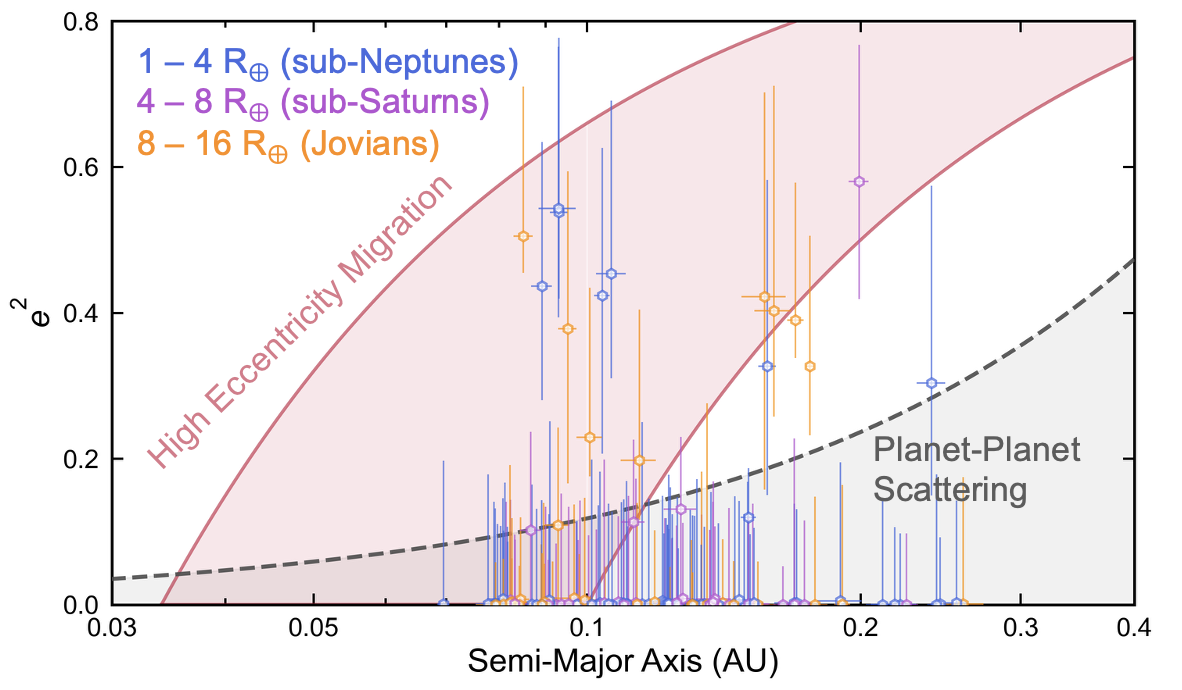}
    \caption{Orbital eccentricity (in $e^2$) versus semi-major axis for the planet sample with eccentricities constrained to better than 40\% for clarity (or if $e<0.1$). Colors denote sub-Neptunes (1--4 $R_\oplus$, blue), sub-Saturns (4--8 $R_\oplus$, purple), and Jovians (8--16 $R_\oplus$, orange). Overlaid curves show theoretical formation channels from \citet{Dawson:2018} for a fiducial planet of $\sim$6 $R_\oplus$ and $\sim$80 $M_\oplus$: high-eccentricity migration (red shaded region) and planet-planet scattering (gray dashed line, gray shaded region). The concentration of larger planets at high eccentricities and small semi-major axes is consistent with high-eccentricity migration.}
    \label{fig:eccplot}
\end{figure}

\begin{figure}
    \includegraphics[width=0.99\textwidth]{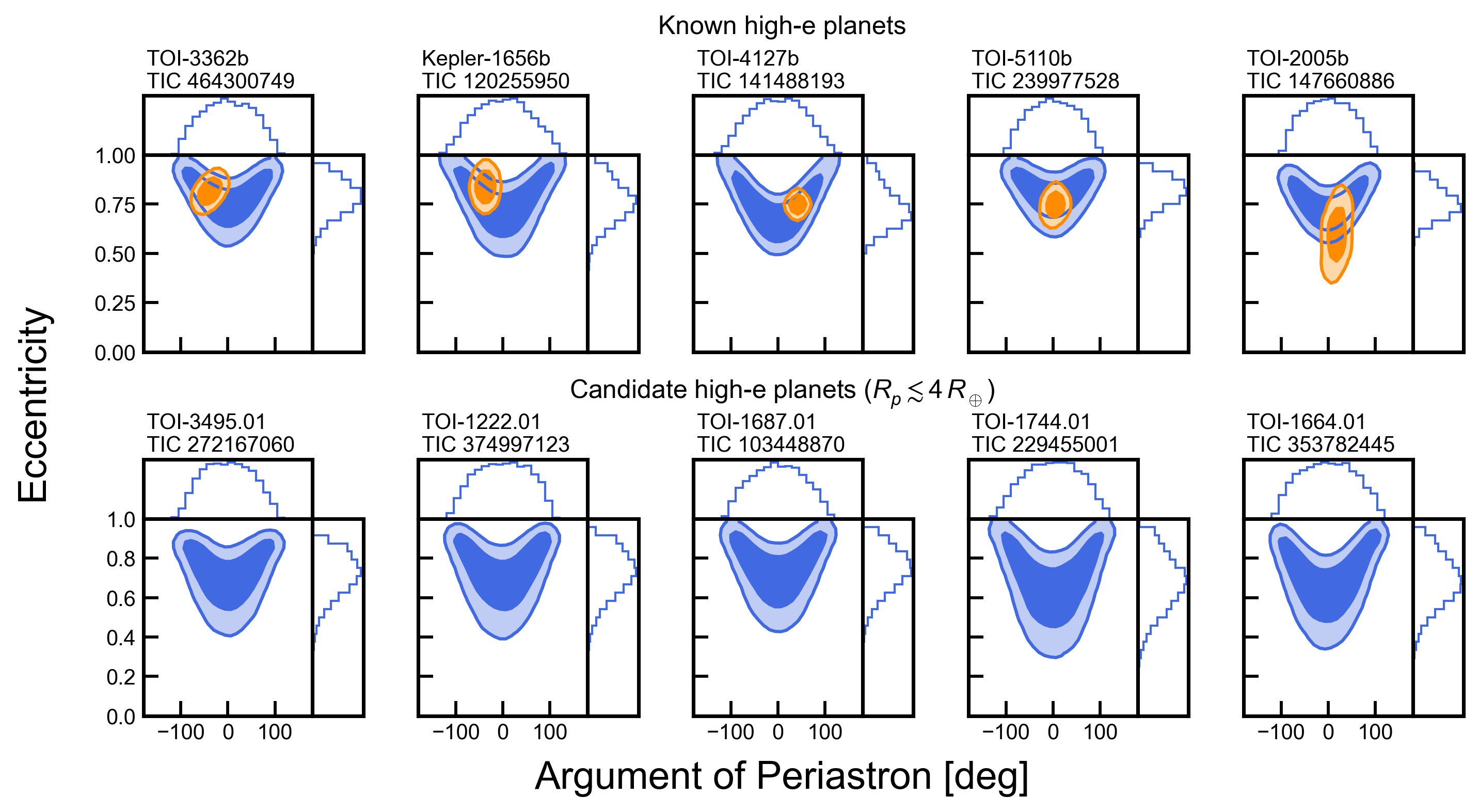}
    \caption{Joint posterior constraints of $e$--$\omega$ from our photoeccentric-effect analysis. Each mini-corner panel shows the marginalized posteriors of $\omega$ (top) and $e$ (right), along with their joint distribution (lower-left). The top row displays previously known high-eccentricity systems, while the bottom row shows newly identified candidate high-$e$  planets with $R_p \lesssim 4,R_\oplus$. For the known systems, the published radial-velocity solutions \citep{toi3362, toi3362_new, Gupta:2023, toi5110, Bieryla:2025} are overlaid as orange Gaussian approximations in $e$ and $\omega$ for visual comparison.}
    \label{fig:known_ecc_comparison}
\end{figure}

In Figure~\ref{fig:eccplot}, we summarize the individual eccentricity constraints across the sample. We report the posterior mode and 68\% credible intervals, since medians can be biased high near the physical boundary at $e=0$. For confirmed planets with published eccentricities, we find agreement within uncertainties; illustrative comparisons are shown in Figure~\ref{fig:known_ecc_comparison}, including new high eccentricity small candidates. We also display the small candidates in Appendix \ref{app:smallplanet} in the context of their host star properties.

\section{Results}
\label{sec:results}
\subsection{Hierarchical inference of the population eccentricity distribution}

Transiting planets dominate the census of confirmed exoplanets due to the scalability of the detection method and dedicated surveys. However, this comes with the compromise that transit light curves constrain eccentricity weakly. Unless the planet is highly eccentric, the uncertainty typically dominates any meaningful constraint in the joint $e-\omega$ space. Yet, while individual measurements may be uncertain, a sample of hundreds of transiting planets allows for the inference of the observed population distribution. Given that warm planets possess circularization timescales long enough to preserve primordial signatures, their eccentricity distribution offers a window into their formation channels.

To this end, we use Hierarchical Bayesian Modeling (HBM) to characterize the observed eccentricity distribution of warm, single planets in \tess. In HBM, we assume that each individual planet is drawn from a parent population governed by common formation channels and dynamical evolution mechanisms. This, in turn, structure allows the pooled data to inform the posteriors of individual objects, shrinking uncertainties as the sample size grows. We follow \citet{Dong:2021} to apply HBM to the eccentricity distribution using the photoeccentric effect.

We assume the eccentricity distribution follows a Beta distribution, chosen for its flexibility and bounded domain on [$0,1$]. Previous literature has demonstrated the Beta distribution sufficiently capture shapes exhibited by planet populations \citep{Kipping_2014b, VanEylen:2019, Dong:2021, Gilbert:2025, Sagear:2025}. We also use a Beta Mixture model (a linear combination of Beta models) to account for multiple sub-populations \citep{Fairnington:2025}, as it is able to better represent asymmetries and tails in eccentricity modes than that of a Gaussian mixture. 

\begin{figure}
    \includegraphics[width=0.99\textwidth]{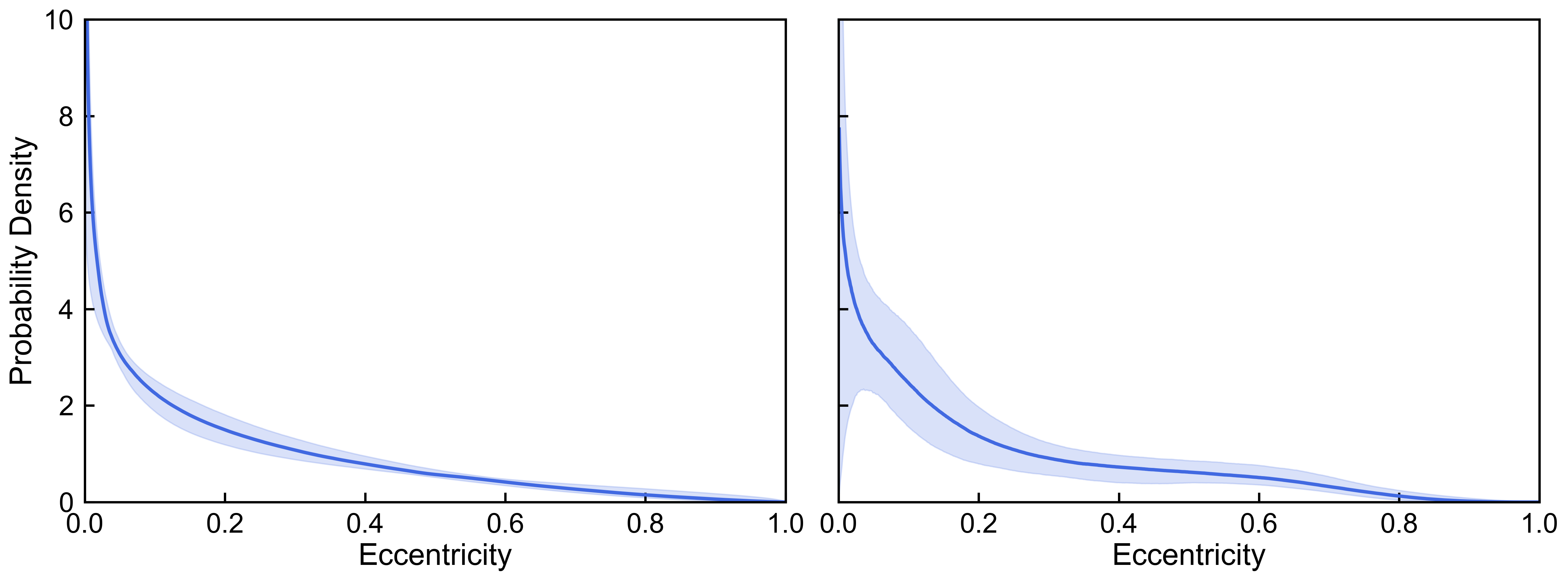}
    \caption{Posterior eccentricity distribution for the full warm-single planet sample (1--16 $R_\oplus$) inferred with our hierarchical Bayesian framework. \textit{Left:} Beta model. \textit{Right:} Two-component Beta mixture model. The solid blue curve shows the median posterior density as a function of eccentricity, and the shaded band indicates the 68\% credible interval.}
    \label{fig:population_distributions}
\end{figure}

We first model the entire transiting sample ($N=219$) under both parametric assumptions. Our hierarchical model follows the structure described in \citet{Fairnington:2025}, in which we infer the re-parameterized mean and concentration of the Beta distribution ($\mu$ and $\kappa$), which deterministically map to the shape parameters $\alpha$ and $\beta$. We approximate the observed parameters $\tilde{\rho}$ and $\rho_\star$ as Gaussian, adding their uncertainties in quadrature. Previous literature has demonstrated that the results are consistent with those using full posterior chains \citep{Dong:2021, Fairnington:2025}. To validate this, we performed a full population analysis using the full posterior chains and found our results are consistent consistent to well within 1$\sigma$. We also importantly reproduce asymmetric posterior eccentricities as expected, as seen in Figure \ref{fig:known_ecc_comparison}.

Because our sample consists exclusively of planets that are observed to transit, the joint distribution of orbital parameters is not that of a random-orientation population. In particular, eccentric configurations have a higher geometric transit probability relative to circular orbits \citep{Burke:2008, Kipping_2014b}. Following the implementation by \citet{Dong:2021}, we incorporate this viewing-geometry effect directly in the likelihood by conditioning on the event that each system transits. For each planet $i$, we include a factor proportional to the geometric transit probability,
\begin{align}
p(\mathrm{obs}_i \mid \rho_{\star,i}, e_i, \omega_i, P_i) \propto 
\left(\frac{R_{\star,i}}{a_i}\right)
\left(\frac{1 + e_i \sin \omega_i}{1 - e_i^2}\right),
\label{eqn:prob}
\end{align}
with $R_{\star,i}/a_i$ written in terms of $\rho_{\star,i}$ and $P_i$ (e.g., \citealt{Winn:2010, Burke:2008, Kipping_2014b}) and is zero at $e_i \ge 1 - R_{\star,i}/a_i$.

We emphasize that this treatment conditions our population inference on the fact that the planets are transiting, and therefore yields an ``observed transiting'' eccentricity distribution that accounts for viewing-geometry bias. It does not, however, correct for survey- and pipeline-dependent selection effects (e.g., detection efficiency, vetting, and sample definition). An intrinsic eccentricity distribution would require explicit modeling of the survey selection function, such as injection--recovery completeness tests, which are yet not well established for \tess. 

We build our model with \texttt{PyMC5}, employing the \texttt{Blackjax} sampling backend \citep{Blackjax} to leverage the significant speedup of \texttt{JAX}-based compilation \citep{JAX}. We found no logical changes in the results between the samplers, with two-to-three orders of magnitude speedup using \texttt{Blackjax}. Four chains, each with 40,000 tuning steps and 10,000 draws, are used in our analysis. Longer chains are required in the hierarchical analysis due to the increased difficulty in sampling with the weighting of Equation \ref{eqn:prob}. We set a target acceptance rate of 0.99 to navigate the complex posterior space.

\begin{figure}
    \includegraphics[width=0.99\textwidth]{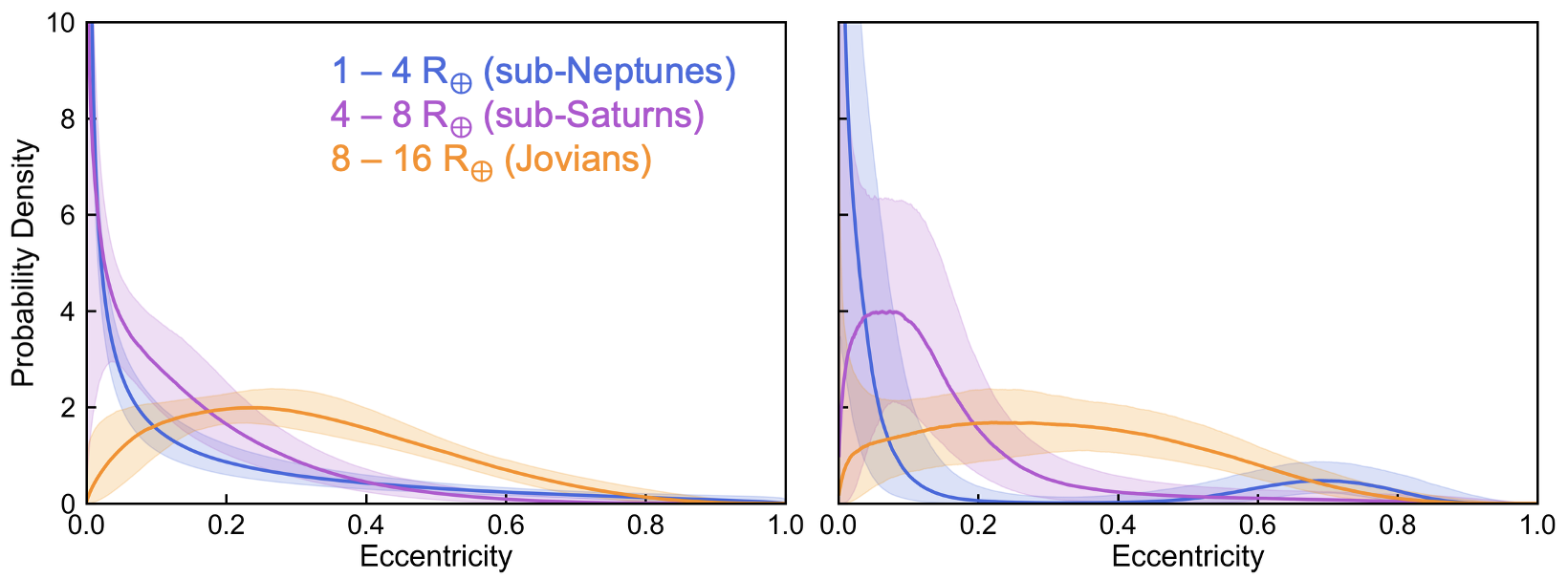}
    \caption{Posterior eccentricity distributions for three planet size classes derived from hierarchical Bayesian inference. \textit{Left:} Single Beta distribution fits. \textit{Right:} Two-component Beta mixture model fits. Colors denote sub-Neptunes (1--4 $R_\oplus$, blue), sub-Saturns (4--8 $R_\oplus$, purple), and Jovians (8--16 $R_\oplus$, orange). Solid lines show median posteriors; shaded regions indicate 68\% credible intervals.}
    \label{fig:beta_distributions}
\end{figure}

\begin{table*}
\centering
\fontsize{6}{8}\selectfont
\caption{Beta and Beta Mixture model results across different planet size ranges}
\label{tab:beta_betamixture_results}
\setlength{\extrarowheight}{2pt}
\renewcommand*{\arraystretch}{1.5}
\begin{tabular}{llllll}
\hline\hline
Distribution & & 1--16 $R_\oplus$ & 1--4 $R_\oplus$ & 4--8 $R_\oplus$ & 8--16 $R_\oplus$ \\
\hline\hline
\multirow{6}{*}{Beta}
 & Fitted:\\
 & $\mu$ & $0.221_{-0.027}^{+0.018}$ & $0.139_{-0.036}^{+0.040}$ & $0.144_{-0.043}^{+0.044}$ & $0.324_{-0.038}^{+0.040}$ \\
 & $\kappa$ & $2.85_{-0.91}^{+0.96}$ & $1.89_{-0.62}^{+0.97}$ & $5.6_{-2.4}^{+4.8}$ & $5.5_{-1.8}^{+2.7}$ \\
 & Derived:\\
 & $\alpha$ & $0.60_{-0.18}^{+0.25}$ & $0.26_{-0.10}^{+0.19}$ & $0.78_{-0.40}^{+0.88}$ & $1.75_{-0.59}^{+0.99}$ \\
 & $\beta$ & $2.25_{-0.74}^{+0.72}$ & $1.63_{-0.53}^{+0.81}$ & $4.8_{-2.0}^{+4.0}$ & $3.7_{-1.2}^{+1.8}$ \\
\multirow{12}{*}{Beta Mixture}
 & Fitted:\\
 & $w_1$ & $0.72_{-0.26}^{+0.16}$ & $0.838_{-0.064}^{+0.052}$ & $0.88_{-0.29}^{+0.08}$ & $0.49_{-0.27}^{+0.33}$ \\
 & $\mu_1$ & $0.118_{-0.070}^{+0.058}$ & $0.025_{-0.018}^{+0.034}$ & $0.106_{-0.049}^{+0.045}$ & $0.20_{-0.13}^{+0.09}$ \\
 & $\kappa_1$ & $10_{-6}^{+20}$ & $21_{-14}^{+37}$ & $20_{-13}^{+32}$ & $11_{-6}^{+21}$ \\
 & $w_2$ & $0.28_{-0.16}^{+0.26}$ & $0.162_{-0.052}^{+0.064}$ & $0.12_{-0.08}^{+0.29}$ & $0.51_{-0.33}^{+0.27}$ \\
 & $\mu_2$ & $0.49_{-0.16}^{+0.15}$ & $0.675_{-0.087}^{+0.080}$ & $0.52_{-0.30}^{+0.26}$ & $0.45_{-0.09}^{+0.13}$ \\
 & $\kappa_2$ & $10_{-6}^{+24}$ & $28_{-16}^{+40}$ & $13_{-8}^{+28}$ & $12_{-6}^{+18}$ \\
 & Derived:\\
 & $\alpha_1$ & $1.0_{-0.5}^{+1.5}$ & $0.4_{-0.4}^{+1.6}$ & $2.0_{-1.4}^{+4.0}$ & $1.8_{-1.1}^{+3.0}$ \\
 & $\beta_1$ & $9_{-5}^{+18}$ & $20_{-13}^{+36}$ & $18_{-12}^{+29}$ & $9_{-5}^{+19}$ \\
 & $\alpha_2$ & $5_{-4}^{+15}$ & $19_{-11}^{+28}$ & $6_{-5}^{+19}$ & $6_{-3}^{+10}$ \\
 & $\beta_2$ & $5.2_{-2.4}^{+8.1}$ & $9_{-5}^{+13}$ & $6_{-3}^{+10}$ & $6.4_{-3.0}^{+7.5}$ \\
\hline
\end{tabular}
\end{table*}
We also performed an archetype-motivated analysis, splitting the population into three distinct classes: sub-Neptunes ($\mathbf{1}-4\ R_{\oplus}$), sub-Saturns ($4-8\ R_{\oplus}$), and Jovians ($8-16\ R_{\oplus}$). While the boundaries of the sub-Saturn population are debated, the compositional diversity in this radius range distinguishes them from their smaller and larger counterparts \citep{Thomas:2025, Petigura:2017}. Additionally, typical radius uncertainties of $\sim 20\%$ imply that planets near a boundary may technically belong to an adjacent population. A further investigation of the radius dependency, with full propagation of radius uncertainty, is described in Section \S\ref{subsec:eccradius}. We display the results of each sub-population's distribution in Figure \ref{fig:beta_distributions}. Additionally, the resulting posterior hyperparameters can be seen in Table \ref{tab:beta_betamixture_results}

\subsection{Hierarchical modeling of the eccentricity--radius relation}
\label{subsec:eccradius}

A primary motivation for this work is to investigate fundamental relationships between eccentricity and planet size. Previous studies have approached this sequentially by modeling discrete radius bins and then fitting a curve to the results. This approach, however, ignores the probability that a planet near a bin edge belongs to a neighboring bin. \citet{Gilbert:2025} demonstrated the importance of propagating radius uncertainties into the $\langle e \rangle - R_p$ model. Here, we develop a 3-stage hierarchical model to self-consistently incorporate radius uncertainty. We simultaneously infer both the size-marginalized population eccentricity distribution and the $\langle e \rangle - R_p$ relation.

Extending the 2-stage HBM from \citet{Dong:2021}, we model the population as a linear combination of two Beta distributions (i.e., a Beta mixture), where the mixture weights are determined by a sigmoid function of radius. The two components correspond to a low-eccentricity and a high-eccentricity population. The mixture weights are governed by a logistic sigmoid function dependent on planet size, such that the probability of belonging to the low-eccentricity component transitions from small to large radii. This allows the data to probabilistically determine the fractional membership of each planet.

We denote with $\pi_{\rm low}(R_{p,i})$ the probability that planet $i$ is drawn from the low-$e$ component (with $\pi_{\rm high}(R_{p,i}) = 1-\pi_{\rm low}(R_{p,i})$ for the high-$e$ component). We define the logistic sigmoid function as
\begin{align}
    \sigma(x)\equiv [1+\exp(-x)]^{-1}
\end{align} 
and write the radius-dependent low-$e$ fraction as
\begin{align}
\pi_{\rm low}(R_{p,i}) &=
\pi_{\rm low,large} + \left(\pi_{\rm low,small}-\pi_{\rm low,large}\right)\,
\sigma\!\left(-\lambda x_i\right),\\
   x_i &\equiv \log R_{p,i} - \log R_{\rm br}\nonumber \\
\label{eq:sigmoid_weight}
\end{align}
where $\pi_{\rm low,small}$ describes the low-$e$ membership fraction for planets well below the break radius ($R_p \ll R_{\rm br}$), $\pi_{\rm low,large}$ describes the same fraction for planets well above the break radius ($R_p \gg R_{\rm br}$), $\lambda$ is the steepness of the transition between the two populations, and $R_{\rm br}$ is the breakpoint radius at which the fractional membership changes.

Each component follows a Beta distribution parameterized by mean $\mu$ and concentration $\kappa$. To ensure identifiability, we impose $\mu_{\rm low} < \mu_{\rm high}$, representing the transition from the low-$e$ to the high-$e$ population. The priors are specified as:
\begin{align}
p(\log R_{p,i}) &\sim \mathcal{N}\!\left(\log \hat{R}_{p,i}, \ \frac{\sigma_{\hat{R}_{p,i}}}{\hat{R}_{p,i}}\right) \nonumber\\
p(\lambda) &\sim \mathcal{U}(0, 30) \nonumber\\
p(\log R_{\rm br}) &\sim \mathcal{U}(\log 2\ R_\oplus, \log 16\ R_\oplus) \nonumber\\
p(\pi_{\rm low,small}, \pi_{\rm low,large}) &\sim \mathcal{U}(0, 1), \quad \pi_{\rm low,small} \ge \pi_{\rm low,large} \nonumber\\
p(\mu_{\rm low}, \mu_{\rm high}) &\sim \mathcal{U}(0, 1), \quad \mu_{\rm low} < \mu_{\rm high} \nonumber\\
p(\log \kappa_{\rm low}, \log \kappa_{\rm high}) &\sim \mathcal{N}(3, 1). \nonumber
\end{align}

This framework captures the physical motivation of two origin channels while maintaining a fully self-consistent treatment of observational uncertainties. The resulting distribution naturally yields the $\langle e \rangle$ vs.\ $R_p$ trend:
\begin{align}
\bar{e} \equiv \mathbb{E}[e \mid R_p] =
\pi_{\rm low}(R_p)\,\mu_{\rm low} + \left[1 - \pi_{\rm low}(R_p)\right]\mu_{\rm high}.
\end{align}
This serves as a probabilistic forecast of eccentricity given radius for the observed warm, single population.

We preferentially define the median eccentricity as a function of planet radius as the 50th percentile of the conditional cumulative distribution function (CDF), which we evaluate numerically by computing the mixture CDF on a dense grid in $e$ and inverting it via linear interpolation. This median is reported in our proceeding results. 

\begin{figure}[h]
    \centering
    \includegraphics[width=0.8\textwidth]{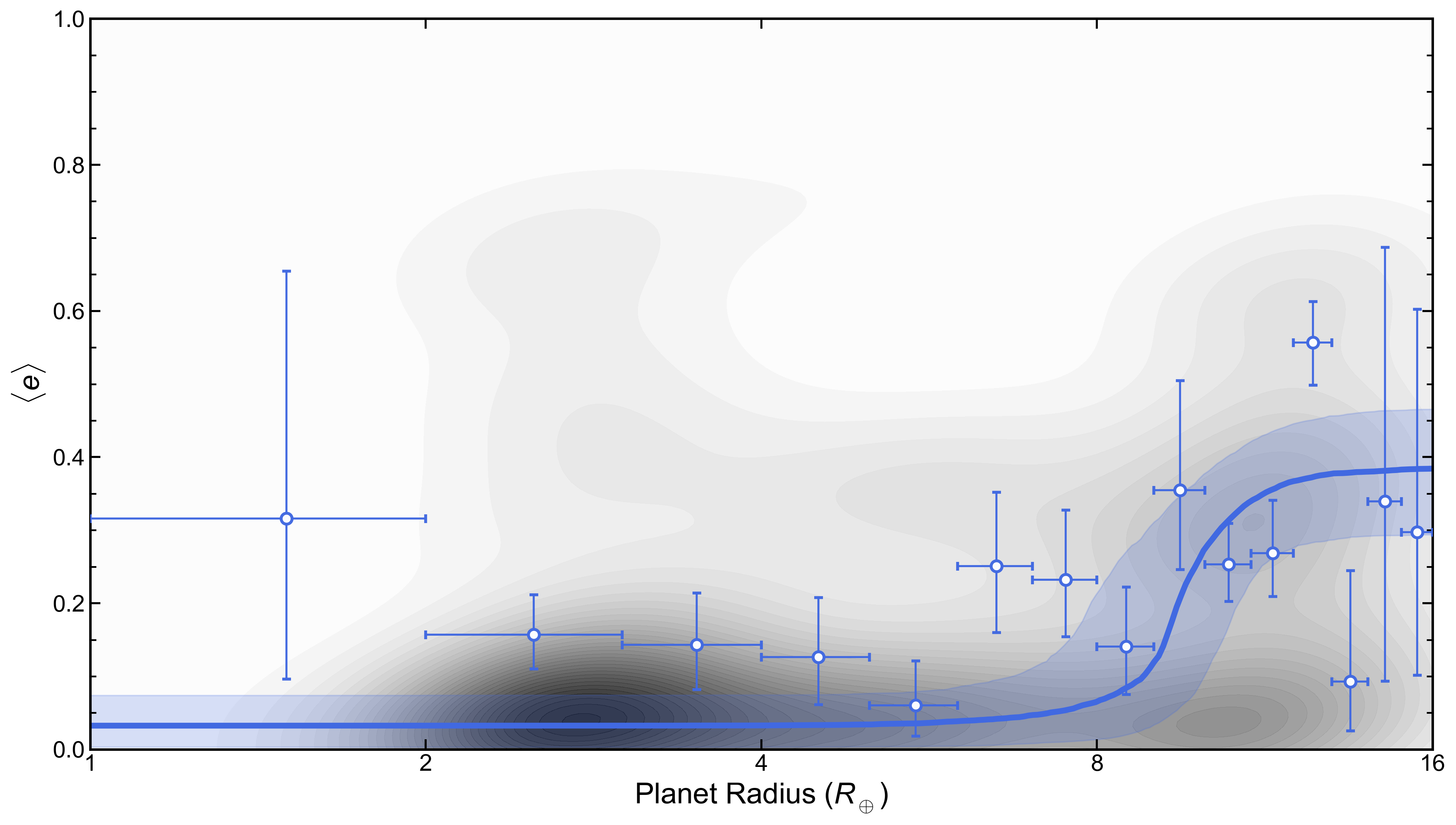}
    \caption{Radius dependence of eccentricity for the warm  sample. Gray contours show the density of the individual-planet eccentricity posterior modes from Section \S\ref{subsec:individualeccentricities}. Open circles with error bars show the radius-binned (``discrete'') analysis, where independent Beta distributions are fit in 1\,$R_\oplus$ bins and summarized by the inferred mean eccentricity in each bin. The solid blue curve and shaded region show the posterior median and 68\% credible interval of the radius-continuous hierarchical model, in which the eccentricity distribution is a two-component Beta mixture with sigmoid-weighted membership as a function of $R_p$.}
    \label{fig:eccentricity_vs_radius}
\end{figure}

We perform two individual analyses, the first being fitting a Beta distribution to the planet size range (1--16 $R_{\oplus}$) with 1\ $R_{\oplus}$ bins, and the latter being our new approach. For the former, we split our population into 15 discrete radius bins encompassing 1-16 $R_\oplus$ and fit a single Beta distribution to each. We run the model with \texttt{PyMC5} again using the \texttt{Blackjax} backend. We use four chains each with 40,000 tuning steps and 10,000 draws, along with a 0.99 target acceptance rate. These are presented as the points in Figure \ref{fig:eccentricity_vs_radius}. For the new hierarchical model, we perform a single analysis over the full size range. This composes the curve in Figure \ref{fig:eccentricity_vs_radius}. The hyperparameter corner plot of the latter analysis can also be found in Appendix \ref{app:app_fig_corner_3stage}. We also demonstrate the effect of binning discretely at individual bin-sizes compared to the continuous model in Appendix \ref{app:bin_sizes}.

To ensure robustness of the retrieved population trend, we validated our results against a highly reliable subset of our sample, namely confirmed planets and/or those cleared by SG1 photometry. We repeated both the individual Beta model inference fits and the 3-stage model, but first kept only confirmed and SG1-cleared planets. Performing this cut resulted in preserving 161 out of the 219 full planet sample, indicating that 58 planets have not received photometric follow-up and have higher intrinsic FP rates. We show the results of this robustness analysis in Appendix~\ref{app:eccentricity_vs_radius_confirmed}. Overall, we find consistent results between our fiducial and confirmed/SG1-cleared sample.

We also performed an independent analysis of the population with F-type stars removed ($>6000$ K) to investigate the dependence on stellar type. We find that the sample without F-type stars remains consistent with the full sample, but we caveat this with the fact that small number statistics (or an absence of giant planets in some bins) could influence these conclusions. This can be seen in Appendix \ref{app:stellar_dependence}.

\section{Discussion}
\label{sec:discussion}

\begin{table*}
\centering
\fontsize{8}{10}\selectfont
\caption{Posterior hyperparameters of the radius-continuous hierarchical model in which the eccentricity distribution is a two-component Beta mixture with a radius-dependent mixture fraction governed by a logistic sigmoid. $R_{\rm br}$ is the break radius and $\lambda$ is the transition steepness; $(\mu_{\rm low},\kappa_{\rm low})$ and $(\mu_{\rm high},\kappa_{\rm high})$ are the mean and precision of the low- and high-$e$ Beta components, respectively; and $\pi_{\rm low,small}$ ($\pi_{\rm low,large}$) is the asymptotic low-$e$ membership fraction for planets with $R_p \ll R_{\rm br}$ ($R_p \gg R_{\rm br}$). The right side reports bin-marginalized low-$e$ fractions $\bar{\pi}_{\rm low}$ and implied mixture-mean eccentricities $\mu_{\rm mix}=\bar{\pi}_{\rm low}\mu_{\rm low}+(1-\bar{\pi}_{\rm low})\mu_{\rm high}$ for three archetype bins. Values are posterior medians with 68\% credible intervals.\label{tab:sigmoid_results}}
\setlength{\extrarowheight}{2pt}
\renewcommand*{\arraystretch}{1.3}
\begin{tabular}{ll @{\hskip 0.4in} ll}
\hline\hline
\multicolumn{2}{c}{\textbf{Sigmoid Mixture Hyperparameters}} & \multicolumn{2}{c}{\textbf{Radius-Marginalized Quantities}} \\
Parameter & Value & Archetype & Value \\
\hline
$R_{\rm br}$ ($R_\oplus$) & \Rbr & \multicolumn{2}{l}{\textit{Sub-Neptunes (1--4 $R_\oplus$)}} \\
$\lambda$                 & \Lam & \hspace{1em}$\bar{\pi}_{\rm low, SN}$ & \wLowSN \\
$\mu_{\rm low}$           & \muOne & \hspace{1em}$\mu_{\rm mix, SN}$ & \muMixSN \\
\cline{3-4}
$\mu_{\rm high}$          & \muTwo & \multicolumn{2}{l}{\textit{Sub-Saturns (4--8 $R_\oplus$)}} \\
$\kappa_{\rm low}$        & \kapOne & \hspace{1em}$\bar{\pi}_{\rm low, SS}$ & \wLowSS \\
$\kappa_{\rm high}$       & \kapTwo & \hspace{1em}$\mu_{\rm mix, SS}$ & \muMixSS \\
\cline{3-4}
$\pi_{\rm low, small}$    & \wmax & \multicolumn{2}{l}{\textit{Jovians (8--16 $R_\oplus$)}} \\
$\pi_{\rm low, large}$    & \wmin & \hspace{1em}$\bar{\pi}_{\rm low, JOV}$ & \wLowJOV \\
& & \hspace{1em}$\mu_{\rm mix, JOV}$ & \muMixJOV \\
\hline
\end{tabular}
\end{table*}

In this work, we characterize the eccentricity distribution of 219 warm, single planets from the \tess{} Prime and Extended missions using the photoeccentric effect \citep{Dawson:2012, Kipping:2010} within a hierarchical Bayesian framework. Because eccentricity is only weakly constrained for most individual transiting planets, the strength of this approach is that modeling the population jointly informs the parent distribution whilst shrinking individual-level uncertainties. Below we interpret the inferred population structure, then discuss the eccentricity--radius transition recovered by our radius-continuous hierarchical model, and finally place these results in the context of prior \tess{} and \kepler{} population studies. 

\subsection{Population-level eccentricity distribution}
\label{ssec:pop_dist}

Under the assumption that the full warm-single population is best described by a single Beta distribution (Figure \ref{fig:population_distributions}, left panel), we infer a distribution that is strongly concentrated around $e=0$ with an extended tail to high eccentricities, with posterior median population mean $\langle\mu\rangle = 0.221_{-0.027}^{+0.018}$ (Table~\ref{tab:beta_betamixture_results}). This result indicates that the typical warm single in our sample is consistent with moderate eccentricity.

However, a single continuous distribution can convolve substructure, and we thus also adopt a two-component Beta mixture model (Figure \ref{fig:population_distributions}, right panel). The population largely still appears consistent with the single Beta model to within 1$\sigma$. If one assumes that the mixture model captures the convolution of two populations, the sample settles on a primarily low-eccentricity component with weight $w_{1} = 0.72_{-0.26}^{+0.16}$ and mean $\mu_1 = 0.118_{-0.070}^{+0.058}$, plus a secondary high-eccentricity component with weight $w_2 = 0.28_{-0.16}^{+0.26}$ and mean $\mu_2 = 0.49_{-0.16}^{+0.15}$ (Table~\ref{tab:beta_betamixture_results}). We find stronger evidence of this population mixture in the next Section. Overall, the low-$e$ component is slightly elevated, but consistent with formation and/or migration pathways that preserve or damp eccentricities (e.g., disk migration or \textit{in-situ} formation, \citealt{Lin:1996, Batygin:2016, Boley:2016}), whereas the high-$e$ component could point to post-disk dynamical excitation, but may also be consistent with migration in a cavity of the protoplanetary disk \citep{Debras:21, Romanova:23}. We note that we do not compute Bayesian evidences for the hierarchical models considered here, so discussion of the mixture model is purely a physically motivated description of the population rather than a formally selected model class.

We also examine archetype-motivated radius classes to compare with expectations in the literature, separating the population into sub-Neptunes (1--4 $R_{\oplus}$), sub-Saturns (4--8 $R_{\oplus}$) and Jovians (8--16 $R_{\oplus}$). In single-Beta fits, both sub-Neptunes and sub-Saturns favor low-to-moderate eccentricities (means $\mu = 0.139_{-0.036}^{+0.040}$ and $\mu = 0.144_{-0.043}^{+0.044}$, respectively), while Jovians show a substantially higher mean $\mu = 0.324_{-0.038}^{+0.040}$ (Table~\ref{tab:beta_betamixture_results}). In the two-component mixture, the Jovian sample also exhibit the largest high-$e$ contribution, with $w_2 = 0.51_{-0.33}^{+0.27}$, qualitatively consistent with previous studies, and the view that warm giants more frequently exhibit signatures of strong dynamical excitation and are formed through multiple mechanisms \citep{Dawson:2018, Dong:2021}. 

\subsection{The eccentricity--radius relationship}
\label{ssec:ecc_rad}

We investigated the eccentricity--radius dependence with two complementary approaches: a radius-binned (``discrete'') analysis and a radius-continuous hierarchical model that propagates radius uncertainties and allows the mixture fraction to vary smoothly with $R_p$ through a logistic sigmoid function. Qualitatively, the recovered trend from the discrete model is that small planets are predominantly drawn from a low-eccentricity component, while the contribution of a high-eccentricity component increases towards larger radii. This narrative is also reinforced through the uniform eccentricity analysis and the resulting contours in Figure~\ref{fig:eccentricity_vs_radius}. The discrete analysis is useful for a first-look visualization, but it is inherently sensitive to bin edges and becomes noisy when the effective number of planets per bin is modest. Moreover, it does not account for the non-negligible probability that planets near a boundary belong to adjacent bins when radius uncertainties are $\sim 20\%$, which can bias inferred trends. Attempting to fit discrete mixture models are even more limited by small-number statistics within each bin, with the number of free parameters sometimes rivaling the number of planets, making such an analysis an overfitting problem. 

The continuous 3-stage hierarchical model alleviates these limitations by using the full sample simultaneously, treating each planet's radius as a latent variable, and inferring a smooth transition in the low-$e$ mixture fraction $\pi_{\rm low}(R_p)$ with radius. We recover a consistent trend to the discrete model, demonstrating that the continuous model recovers the same qualitative behavior while reducing binning-induced noise. In this model, the component means are $\mu_{\rm low}=$ \muOne{} and $\mu_{\rm high}=$ \muTwo{} (Table~\ref{tab:sigmoid_results}). The transition in mixture fraction occurs at a break radius $R_{\rm br} =$  \Rbr{} $R_\oplus$ with steepness $\lambda =$ \Lam{} (Table~\ref{tab:sigmoid_results}). 

As the components are mathematical limits, to report quantitatively useful results, we marginalize over aforementioned archetype size definitions and compute the bin-averaged low-$e$ fraction $\bar{\pi}_{\rm low}([R_1,R_2])$ directly from posterior draws. For sub-Neptunes (1--4 $R_\oplus$) we infer $\bar{\pi}_{\rm low}=\wLowSN$ (corresponding to $\bar{\pi}_{\rm high}=\wHighSN$), for sub-Saturns (4--8 $R_\oplus$) $\bar{\pi}_{\rm low}=\wLowSS$ ($\bar{\pi}_{\rm high}=\wHighSS$), and for Jovians (8--16 $R_\oplus$) $\bar{\pi}_{\rm low}=\wLowJOV$ ($\bar{\pi}_{\rm high}=\wHighJOV$). These are the quantities summarized by the colored points in Figure~\ref{fig:weight_vs_rp}.

Overall, the model indicates that a non-zero high-eccentricity component may be present across the full radius range, with its relative contribution increasing toward larger planets through the decline of $\pi_{\rm low}(R_p)$ with $R_p$.

\subsection{Implications for formation and dynamical evolution across planet sizes}
\label{ssec:formation}

\begin{figure}[h]
\centering
\includegraphics[width=0.49\textwidth, keepaspectratio]{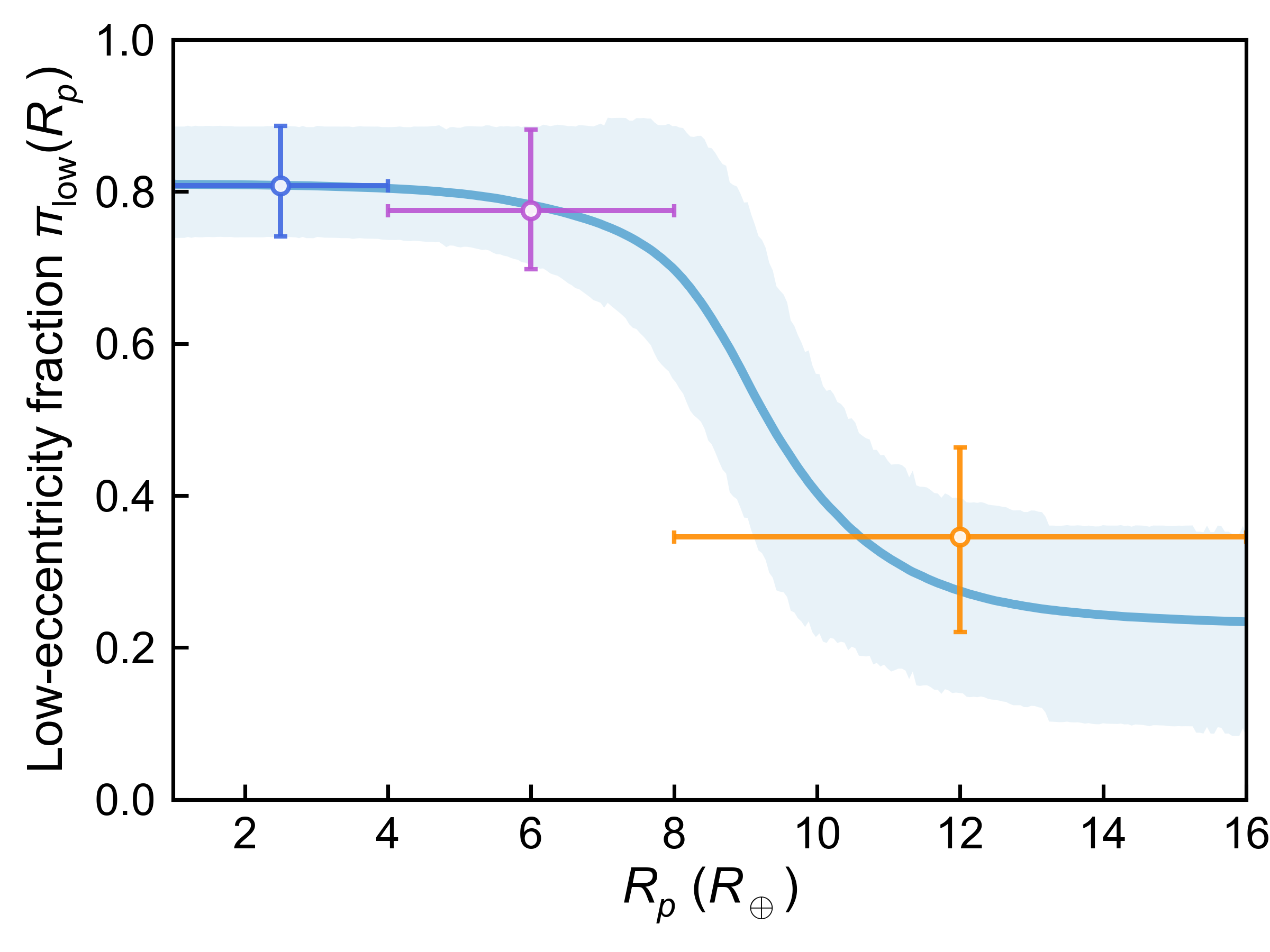}
\caption{Inferred low-eccentricity mixture fraction $\pi_{\rm low}(R_p)$ from the continuous sigmoid mixture model of the final sample. The line and shaded band denote the posterior median and 68\% credible interval. Coloured points summarize the corresponding size-marginalized low-$e$ fractions in three archetype radius bins (1--4, 4--8, and 8--16 $R_\oplus$), with horizontal bars showing bin extents and vertical bars showing 68\% credible intervals.}
\label{fig:weight_vs_rp}
\end{figure}

The inferred transition at $R_{\rm br}\approx 9\,R_\oplus$ suggests that warm planets below and above this size are typically shaped by different dynamical histories. Sub-Neptunes and sub-Saturns are primarily assigned to the low-$e$ component, with size-marginalized low-$e$ fractions of $\bar{\pi}_{\rm low}=\wLowSN$ for 1--4 $R_\oplus$ and $\bar{\pi}_{\rm low}=\wLowSS$ for 4--8 $R_\oplus$ (Figure~\ref{fig:weight_vs_rp}). This is consistent with formation and/or migration pathways that do not typically produce large eccentricities, or with efficient damping during the gas-disk phase \citep{Goldreich:1980, Bitsch:2013, Duffell:2015}. The dynamical similarity between sub-Saturns and sub-Neptunes in our warm-single sample is notable given the compositional diversity often discussed for 4--8 $R_\oplus$ planets \citep{Thomas:2025, Petigura:2017}. Nevertheless, our results also suggest a non-zero population of eccentric small planets (high-$e$ fractions $\bar{\pi}_{\rm high}=\wHighSN$ and $\wHighSS$), which should be of particular intrigue for future follow-up of the target list.

In contrast, warm giants show a substantially larger inferred contribution from the high-$e$ component. Marginalizing over the Jovian regime (8--16 $R_\oplus$), we infer a high-$e$ fraction of $\bar{\pi}_{\rm high}=\wHighJOV$ (Figure~\ref{fig:weight_vs_rp}). 
Disk-free migration channels such as planet-planet scattering and secular perturbations from external companions, such as the Kozai-Lidov mechanism and secular chaos, produce warm giants with elevated eccentricities \citep{Rasio&Ford:1996, Chatterjee:2008, Wu:2011, Naoz:2016}. Coplanar high-eccentricity migration has also been proposed as a route to high $e$ without necessarily generating large mutual inclinations \citep{Petrovich:2015a}, and can be tested with individual-level constraints in follow-up. The latter mechanism has also gained favor due to recent findings of predominantly aligned eccentric systems \citep{Rice:2022, Bieryla:2025}. 

\begin{figure}
\centering
    \includegraphics[width=0.99\textwidth]{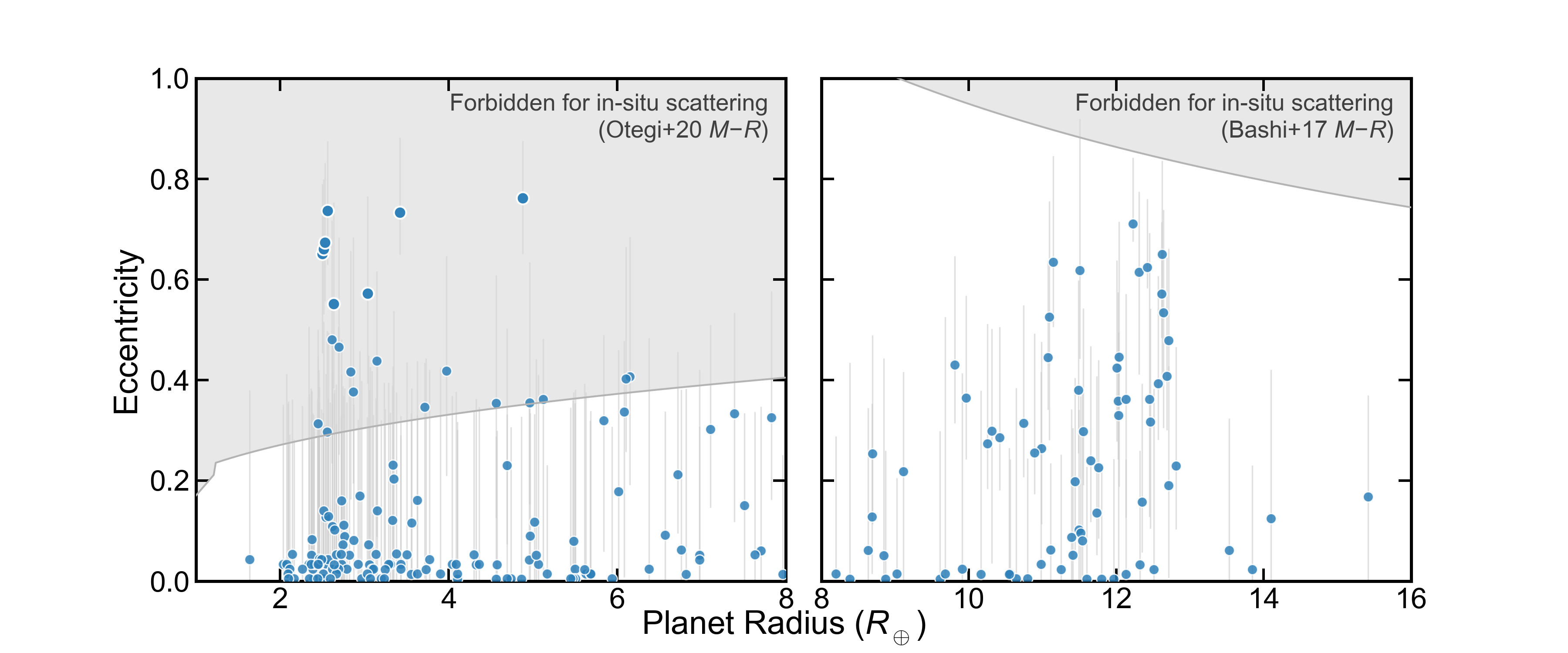}
    \caption{Eccentricity--radius diagram for the planet sample. Points show the mode eccentricity, with errors drawn from 68\% credible intervals. The curves indicate first-order scattering-based characteristic eccentricities, \(e_{\rm sc}=\sqrt{\Theta}\), evaluated at fixed ($a=0.2$) AU and ($M_\star=1\ M_\odot$), where ($\Theta=2(a/R_p)(M_p/M_\star)$). Two mass-radius assumptions bound the maximum eccentricity curve: the empirical \citet{Otegi:2020} relation for planets below $120\ M_\oplus$, which we cut off at our canonical sub-Saturn size, dashed line; and a Jupiter-like giant-planet scaling from \citet{Bashi:2017}, solid line. These composition-dependent curves provide illustrative upper-envelope expectations for eccentricities produced by planet-planet scattering.
}
    \label{fig:compositions}
\end{figure}

The inferred eccentricities of warm single giants in our analysis is equally consistent, however, with a population shaped by disk migration alone, whereby some fraction of Jovians undergo in-situ planet-planet scattering after the disk disappears. The maximum eccentricity that planet-planet scattering can create in-situ can be approximated as \citep{Goldreich:2004}
\begin{align}
    e_{\rm sc} = \sqrt{\Theta}
\end{align} 
where $\Theta$ is the Safronov number, defined as
\begin{align}
     \Theta = \frac{v_{\rm esc}^2}{v_{\rm orb}^2} = 2 \frac{M_p}{M_\star}\frac{a}{R_p}
\end{align}
and where $v_{\rm esc}$ is the escape velocity of the planet, $v_{\rm orb}$ is its orbital velocity assuming a circular orbit, $M_p$ is the planet's mass, $M_\star$ is the mass of the star, $a$ is the semi-major axis and $R_p$ is the radius of the planet. For a typical warm Jupiter ($M_p \approx 300\ M_\oplus$, $R_p \approx 10\ R_\oplus$, $a = 0.2$  AU) around a Sun-like star ($M_\star = 1\ M_\odot$), this maximum eccentricity is of order unity, suggesting a Jovian-like planet can reach eccentricities high enough to trigger tidal migration via in-situ scattering, without the need to invoke more complex mechanisms. In Figure \ref{fig:compositions}, we show theoretical $e_{\rm sc}$ upper limits under the assumption of empirical mass-radius relationships \citep{Otegi:2020, Bashi:2017} for both the small (1--8 $R_\oplus$) and giant (8--16 $R_\oplus$) planets. 
This formation scenario also appears consistent with the mass-eccentricity trend and spin-orbit alignments of highly eccentric warm Jupiters \citep{Dong:2026}. The giant planet sample as a function of various planet/stellar properties can be seen in Appendix \ref{app:giantplanet}. 
Recent radial-velocity analyses of warm Jupiters provide an independent comparison point for our inferred high-eccentricity Jovian component. \citet{Morgan:2025} presented a homogeneous catalog of Keplerian fits for 200 warm Jupiters, and \citet{Morgan:2026} used this sample to infer a broad population-level eccentricity distribution in which only $27^{+3}_{-4}\%$ of warm Jupiters are consistent with near-circular orbits ($e<0.1$), while $73^{+3}_{-3}\%$ are dynamically excited. Although their Doppler-selected sample spans longer orbital periods ($\approx10$--365 days) and is not directly comparable to our transiting \tess{} warm-single sample ($8<P<50$ days), the prevalence of excited eccentricities among RV warm Jupiters is qualitatively consistent with our finding that the high-$e$ component becomes most prominent in the Jovian-radius regime. This agreement across independent samples and eccentricity-measurement methods supports the interpretation that warm giants commonly experience post-disk dynamical excitation, potentially through scattering or secular interactions.

Finally, the placement of $R_{\rm br}$ above the canonical sub-Saturn regime implies that most sub-Saturns reside on the same low-$e$ track as sub-Neptunes in the warm-single population. This result disfavors a naive expectation that sub-Saturns should systematically share the dynamical signatures of giants if they were predominantly ``failed gas giants'' in the sense of being outcomes of a truncated runaway-accretion phase \citep{Lee:2016}, since dynamical excitation is often linked to giant planet interactions and scattering \citep{Chatterjee:2008, Ford:2008, Nagasawa:2011}. Instead, our inference is consistent with scenarios in which a large fraction of sub-Saturns represent an extension of sub-Neptune formation pathways and/or have architectures that experience fewer strong post-disk interactions.

\subsection{A sub-population of eccentric small planets}
\label{ssec:small_ecc}

Although small planets are dominated by the low-$e$ component, the radius-continuous mixture model assigns a non-zero probability that a small subset of sub-Neptunes belong to the dynamically excited component. Marginalizing over $1$--$4\,R_\oplus$, we infer a high-$e$ fraction of $\bar{\pi}_{\rm high}=\wHighSN$, comparable to (though not identical with) the $4$--$8\,R_\oplus$ bin value of $\bar{\pi}_{\rm high}=\wHighSS$ (Figure~\ref{fig:weight_vs_rp}). To aid follow-up, Table~\ref{tab:small_high_e_targets} summarizes targets with $R_p<6\,R_\oplus$ whose eccentricity posterior mode exceeds $0.5$.

\begin{deluxetable*}{lccccccc}
\tabletypesize{\scriptsize}
\setlength{\tabcolsep}{3pt}
\tablecaption{Targets with $R_p<6\,R_\oplus$ and eccentricity posterior mode $>0.5$. Follow-up is grouped by TFOP SG1 (seeing-limited photometry), SG2 (spectroscopy), and SG3 (high-resolution imaging).\label{tab:small_high_e_targets}}
\tablewidth{0pt}
\tablehead{
\colhead{Target} &
\colhead{$T_{\rm eff}$ (K)} &
\colhead{$R_\star$ ($R_\odot$)} &
\colhead{$R_p$ ($R_\oplus$)} &
\colhead{$e$} &
\colhead{SG1} &
\colhead{SG2} &
\colhead{SG3}
}
\startdata
Kepler-1656b (TOI-4584.01) & $\sim5900$ & $\sim1.00$ & $4.88^{+0.28}_{-0.27}$ & $0.762^{+0.115}_{-0.111}$ & --- & RV (lit.) & --- \\
TOI-3495.01 & $\sim6400$ & $\sim1.46$ & $4.10^{+0.32}_{-0.32}$ & $0.743^{+0.106}_{-0.119}$ & LCO-CTIO & SMARTS/CHIRON & SOAR/HRCam \\
TOI-1222.01 & $\sim6000$ & $\sim1.33$ & $2.56^{+0.16}_{-0.19}$ & $0.737^{+0.139}_{-0.108}$ &
\shortstack[l]{LCO-SSO;\\ PEST} &
SMARTS/CHIRON &
\shortstack[l]{Gemini/Zorro;\\ SOAR/HRCam} \\
TOI-1687.01 & $\sim5970$ & $\sim0.96$ & $3.42^{+0.23}_{-0.26}$ & $0.733^{+0.149}_{-0.084}$ &
\shortstack[l]{LCO (McD/CTIO/Teide);\\ KeplerCam;\\ MuSCAT3} &
FLWO/TRES &
\shortstack[l]{WIYN/NESSI; Gemini/Alopeke;\\ CAHA/AstraLux; SAI} \\
TOI-1744.01 & $\sim6010$ & $\sim1.25$ & $2.97^{+0.24}_{-0.32}$ & $0.690^{+0.138}_{-0.167}$ &
--- &
\shortstack[l]{NOT/FIES; APF/Levy;\\ Keck/HIRES} &
\shortstack[l]{Keck/NIRC2;\\ Gemini/Alopeke} \\
TOI-1664.01 & $\sim6160$ & $\sim1.34$ & $2.53^{+0.14}_{-0.16}$ & $0.674^{+0.158}_{-0.117}$ &
KeplerCam &
\shortstack[l]{FLWO/TRES; Keck/HIRES;\\ LCO/NRES} &
\shortstack[l]{WIYN/NESSI; Shane/ShARCS;\\ CAHA/AstraLux; SAI} \\
HIP~97166b (TOI-1255.01) & $\sim5390$ & $\sim0.84$ & $2.52^{+0.15}_{-0.15}$ & $0.661^{+0.138}_{-0.130}$ & --- & RV (lit.) & --- \\
TOI-5789.01 & $\sim5640$ & $\sim0.82$ & $2.50^{+0.19}_{-0.20}$ & $0.651^{+0.140}_{-0.198}$ &
LCO-McD &
\shortstack[l]{Palomar/DBSP;\\ FLWO/TRES} &
\shortstack[l]{Palomar/PHARO;\\ SOAR/HRCam; SAI} \\
TOI-2100.01 & $\sim5510$ & $\sim0.86$ & $3.04^{+0.23}_{-0.26}$ & $0.572^{+0.193}_{-0.179}$ &
LCO-McD &
\shortstack[l]{FLWO/TRES;\\ Keck/HIRES} &
\shortstack[l]{CAHA/AstraLux; Gemini/Alopeke;\\ Keck/NIRC2; SAI} \\
TOI-1209.01 & $\sim6230$ & $\sim1.34$ & $2.63^{+0.17}_{-0.18}$ & $0.551^{+0.202}_{-0.167}$ &
LCO-CTIO &
SMARTS/CHIRON &
\shortstack[l]{Gemini/Zorro;\\ SOAR/HRCam} \\
\enddata
\tablenotetext{}{A more detailed and precise table is available in machine-readable form.}
\end{deluxetable*}

Two targets in this criteria range have been previously confirmed. Kepler-1656b is a warm sub-Saturn \citep{Brady:2018, angelo:2022}. We infer $e = 0.762^{+0.115}_{-0.111}$ (Figure~\ref{fig:known_ecc_comparison}), consistent with the published RV eccentricity of $e = 0.838^{+0.045}_{-0.029}$ \citep{angelo:2022}. HIP~97166b (TOI-1255.01) illustrates a known limitation of photoeccentric constraints. While we infer $e = 0.661^{+0.138}_{-0.130}$, an RV analysis finds $e = 0.29\pm0.072$ \citep{Polanski:2024}. The joint photometric--RV solution indicates a high impact parameter ($b=0.836\pm0.027$), which can mimic a shortened duration and is partially degenerate with $(e,\omega)$ \citep{Macdougall:2021}.

A straightforward interpretation of the remaining cases is dynamical excitation by additional companions (e.g., secular forcing or scattering). This scenario predicts that a substantial fraction should host massive outer companions detectable via RV trends and/or transit-timing variations. We note, however, that the false-positive probability is likely higher for this subset, reinforcing the importance of continued follow-up.

\subsection{Comparison with previous work}
\label{ssec:comparison}

This work provides the first homogeneous, all-sky characterization of the eccentricity distribution of warm ($8<P<50$~days)  planets across the full \tess{} radius range. By modeling the population with a radius-dependent mixture distribution, we show that the observed warm-planet sample is well described by a low-eccentricity component plus a dynamically excited high-$e$ component whose contribution increases strongly with planet size. This framework unifies previous \kepler{} and \tess{} results into a single picture in which the fraction of high-eccentricity systems rises from small planets to giant planets.

Studies of  systems have consistently found eccentricity distributions that are inconsistent with a population shaped purely by disk-driven formation and evolution. Using a large \kepler{}--LAMOST sample with spectroscopically derived stellar densities, \citet{Xie:2016} showed that  planets exhibit moderate eccentricities, with a population mean of $\langle e \rangle \sim 0.3$. Improved stellar density determination through astroseismology lead to more precise measurements of small-planet eccentricities, in which it was shown that small singles ($R_p < 4\,R_\oplus$) exhibit a reduced pile-up at near-circular orbits and a broader eccentricity distribution relative to multi-transiting systems \citep{VanEylen:2019}. Together, these studies established that elevated eccentricities are a generic feature of  populations across a wide range of planet sizes. We reinforce and extend these conclusions using a homogeneous, all-sky \tess{} sample and precise stellar densities from Gaia.
We also note that several \kepler{}-based samples include planets at shorter orbital periods (no period cutoff in the former, and $P>5$~days in the latter), where tidal circularization is more effective. Including these shorter-period planets can lower the inferred eccentricity scale relative to the warm-planet regime ($8<P<50$~days) studied here.

The growth of the \tess{} sample has enabled targeted studies of warm planets over longer orbital periods. Focusing on warm Jupiters ($R_p \gtrsim 6\,R_\oplus$, $8<P<200$ days), \citet{Dong:2021} inferred a bimodal eccentricity distribution with an approximately 55/45 split between low- and high-eccentricity components. While their analysis did not distinguish between single- and multi-transiting systems, the inferred high-eccentricity fraction is fully consistent with the Jovian high-$e$ fraction we recover, $\bar{\pi}_{\rm high}=\wHighJOV$. Extending this work to smaller radii, \citet{Fairnington:2025} analyzed warm sub-Saturns ($4$--$8\,R_\oplus$, $8<P<200$ days) and found a similarly elevated eccentricity distribution for  systems, including evidence for a $\sim$15\% high-eccentricity sub-population. Our results reproduce this behavior and demonstrate that the emergence of a high-$e$ component continues into the sub-Neptune regime. Collectively, these \tess{}-based studies point to a population in which increasing planet size is accompanied by an increasing contribution from dynamically excited systems, rather than a uniform shift in eccentricity across all planets.

A qualitatively similar size-dependent trend was also reported for \kepler{}  planets by \citet{Gilbert:2025}, who found that eccentricity increases with planet radius and inferred a transition near $\sim3.5\,R_\oplus$. The smaller transition radius in that work likely reflects differences in both orbital-period coverage ($1<P<100$~days) and the higher number of intrinsically rare planets that \tess{} samples. In particular, \citet{Gilbert:2025} aim to recover the intrinsic eccentricity distribution after accounting for geometric and detection selection effects, whereas our analysis (and the other studies discussed above) characterizes the observed transiting population. Because transiting samples are weighted toward configurations with higher transit probability, which can be enhanced for dynamically excited orbits, the observed warm planet population probed by \tess{} may exhibit a higher apparent contribution from excited systems. In this sense, the larger transition radius we infer ($R_{\rm br}\approx9\,R_\oplus$) and the lower-radius transition inferred by \citet{Gilbert:2025} may represent complementary views of the same underlying population rather than a direct discrepancy.

\subsection{Limitations and future work}
\label{ssec:limitations}

Our constraints rely on the photoeccentric effect, which infers eccentricity primarily from transit durations and stellar densities \citep{Dawson:2012, Kipping_2014b}. Our analysis corrects for the geometric over-representation of eccentric transiting orbits, but does not yet account for survey completeness and vetting selection as functions of period, radius, and host properties; a fully intrinsic distribution will require injection–recovery style forward modeling. We also do not perform a dedicated end-to-end injection-and-recovery test in which synthetic systems with a known eccentricity–radius relation are passed through the full light-curve fitting and hierarchical inference pipeline. While such a test would provide an additional validation of the framework, we consider it beyond the scope of the present work. We instead note that our methodology recovers published parameters and eccentricity constraints for confirmed systems in the sample (Figure \ref{fig:known_ecc_comparison}), which provides empirical support that the framework is performing as expected. In addition, we have removed targets with ambiguous orbital periods derived solely from \tess, which may inhibit the representation of long-period targets from our initial sample, especially if they have been confirmed via ground-based follow-up. This was necessary for a homogeneous analysis, but may remove targets of interest (e.g., TOI-2134c, \citealt{Rescigno:2024}).

We note that this analysis makes use of both confirmed planets and current planet candidates. Despite performing extensive vetting to remove targets known to be false positives, not all of our sample are classified as \textit{bona fide} planets. Follow-up of current candidates is therefore of high scientific value.

Although we attempt to mitigate systematics by deriving independent stellar densities from \textit{Gaia} DR3 and SED fitting, unresolved binaries and blended photometry can bias densities and therefore photoeccentric constraints \citep{Furlan:2020}. High-resolution imaging and spectroscopic vetting for subsets of the sample would help further increase the sample's robustness. 

Finally, we emphasize that our focus on  warm planets is a deliberate choice that complements studies that combine singles and multis. Because dynamical stability constraints tend to favor low eccentricities in tightly packed multi-planet systems, contrasting warm singles against warm multis within a consistent framework can isolate dynamical excitation associated with lower-multiplicity architectures \citep{VanEylen:2019, Gilbert:2025}. Extending our analysis to warm multi-transiting systems in \tess{} is therefore a natural next step.

\section{Conclusions}

We have inferred the \tess{} warm-single ($P=8$--50~days) eccentricity distribution with a sample of 219 planets spanning $R_p\sim1$--$16\,R_\oplus$ using the photoeccentric effect within a hierarchical Bayesian framework. Our main findings are:
\begin{itemize}
    \item The warm-single population is well described by both a single Beta distribution peaking at zero and falling monotonically, and a two-component Beta mixture with a dominant low-eccentricity component ($\langle e_{\rm low}\rangle = \muOne$) and a secondary dynamically excited component with mean $\langle e_{\rm high} \rangle=\muTwo$.
    
    \item The contribution of the high-$e$ component increases strongly with planet radius. In our radius-continuous mixture model, the low-$e$ membership fraction transitions with a break radius of $R_{\rm br}=\Rbr\,R_\oplus$, implying that dynamical excitation becomes increasingly common toward warm Jovians.
    
    \item In archetype-motivated radius classes, sub-Neptunes and sub-Saturns are predominantly consistent with the low-$e$ component, while Jovians exhibit the clearest high-$e$ contribution, consistent with warm giants tracing a wider range of eccentricities and more dynamical histories.
    
    \item Although small planets are dominated by low eccentricities, the model assigns a non-zero high-$e$ fraction among sub-Neptunes ($\bar{\pi}_{\rm high}=\wHighSN$ over $1$--$4\,R_\oplus$; similarly $\bar{\pi}_{\rm high}=\wHighSS$ for $4$--$8\,R_\oplus$). We identify a set of small, high-$e$ candidates that are especially valuable targets for confirmation and mass determination.
\end{itemize}

These results support a picture in which warm singles comprise a mixture of relatively quiescent systems and a dynamically excited sub-population, with the latter becoming increasingly dominant at larger radii. Moving from the observed to the intrinsic eccentricity distribution will require forward modeling of survey completeness and vetting selection (e.g., injection-recovery), and an immediate next step is extending this framework to warm multi-transiting \tess{} systems to isolate eccentricity differences linked to planet multiplicity and low mutual inclinations.

\begin{acknowledgments}

We thank the anonymous reviewer for their thoughtful suggestions and comments which have significantly improved the quality of the manuscript. 

We respectfully acknowledge the traditional custodians of the lands on which we conducted this research and throughout Australia. We recognize their continued cultural and spiritual connection to the land, waterways, cosmos, and community. We pay our deepest respects to all Elders, present and emerging, and the people of the Giabal, Jarowair, and Kambuwal nations, upon whose lands this research was conducted.

This material is based upon work supported by the National Aeronautics and Space Administration under Agreement No.\ 80NSSC21K0593 for the program ``Alien Earths''.
The results reported herein benefited from collaborations and/or information exchange within NASA’s Nexus for Exoplanet System Science (NExSS) research coordination network sponsored by NASA’s Science Mission Directorate.
This material is based upon work supported by the European Research Council (ERC) Synergy Grant under the European Union’s Horizon 2020 research and innovation program (grant No.\ 101118581---project REVEAL).

C.X.H acknowledges that her research is sponsored by the Australian Research Council Future Fellowship FT240100016. G.Z. acknowledges that his research is sponsored by the Australian Research Council.

FPW acknowledges support from the Union College Faculty Research Fund. 

Funding for KB was provided by the European Union (ERC AdG SUBSTELLAR, GA 101054354).

We acknowledge financial support from the Agencia Estatal de Investigaci\'on of the Ministerio de Ciencia e Innovaci\'on MCIN/AEI/10.13039/501100011033 and the ERDF “A way of making Europe” through projects PID2021-125627OB-C32 and PID2024-158486OB-C32, and from the Centre of Excellence “Severo Ochoa” award to the Instituto de Astrofisica de Canarias. 
J.L.-B. is funded by the Spanish grants PID2023-150468NB-I00 and CNS2023-144309 from the Ministry of Science, Innovation and Universities (MCIN/AEI/10.13039/501100011033)
F. M. acknowledges the financial support from the Agencia Estatal de Investigaci\'{o}n del Ministerio de Ciencia, Innovaci\'{o}n y Universidades (MCIU/AEI) through grant PID2023-152906NA-I00.
Funding for KB was provided by the European Union (ERC AdG SUBSTELLAR, GA 101054354).
This article is based on observations made with the MuSCAT2 instrument, developed by ABC, at Telescopio Carlos Sánchez operated on the island of Tenerife by the IAC in the Spanish Observatorio del Teide.
This paper is based on observations made with the MuSCAT3/4 instruments, developed by the Astrobiology Center (ABC) in Japan, the University of Tokyo, and Las Cumbres Observatory (LCOGT). MuSCAT3 was developed with financial support by JSPS KAKENHI (JP18H05439) and JST PRESTO (JPMJPR1775), and is located at the Faulkes Telescope North on Maui, HI (USA), operated by LCOGT. MuSCAT4 was developed with financial support provided by the Heising-Simons Foundation (grant 2022-3611), JST grant number JPMJCR1761, and the ABC in Japan, and is located at the Faulkes Telescope South at Siding Spring Observatory (Australia), operated by LCOGT.
This work is partly supported by JSPS KAKENHI Grant Numbers JP24H00017, JP24K00689, JSPS Grant-in-Aid for JSPS Fellows Grant Number JP25KJ0091 and JSPS Bilateral Program Number JPJSBP120249910.


This work makes use of observations from the LCOGT network. Part of the LCOGT telescope time was granted by NOIRLab through the Mid-Scale Innovations Program (MSIP). MSIP is funded by NSF.


This paper is based on observations made with the Las Cumbres Observatory’s education network telescopes that were upgraded through generous support from the Gordon and Betty Moore Foundation.

We acknowledge the use of public TESS data from pipelines at the TESS Science Office and at the TESS Science Processing Operations Center.

This paper includes data collected by the TESS mission that are publicly available from the Mikulski Archive for Space Telescopes (MAST).


This research has made use of the Exoplanet Follow-up Observation Program (ExoFOP; DOI: 10.26134/ExoFOP5) website, which is operated by the California Institute of Technology, under contract with the National Aeronautics and Space Administration under the Exoplanet Exploration Program.


Funding for the TESS mission is provided by NASA's Science Mission Directorate. KAC acknowledges support from the TESS mission via subaward s3449 from MIT.

This research made use of \texttt{exoplanet} \citep{exoplanet:joss,
exoplanet:zenodo} and its dependencies \citep{celerite2,
exoplanet:agol20, exoplanet:arviz,
exoplanet:astropy13, exoplanet:astropy18, exoplanet:kipping13,
exoplanet:luger18}.

This work has made use of data from the European Space Agency (ESA) mission Gaia (https://www.cosmos.esa.int/gaia), processed by the Gaia Data Processing and Analysis Consortium (DPAC, https://www.cosmos.esa.int/web/gaia/dpac/consortium).
\end{acknowledgments}

\facilities{TESS, SOAR, WIYN, LCO, SMARTS, PEST, FLWO, Gemini, CAHA, SAI, NOT, APF, Keck, Shane, Palomar, MuSCAT3}

\appendix
\renewcommand{\figurename}{Appendix}
\renewcommand{\thefigure}{\arabic{figure}}
\setcounter{figure}{0}

\begin{figure*}
    \includegraphics[width=0.99\textwidth]{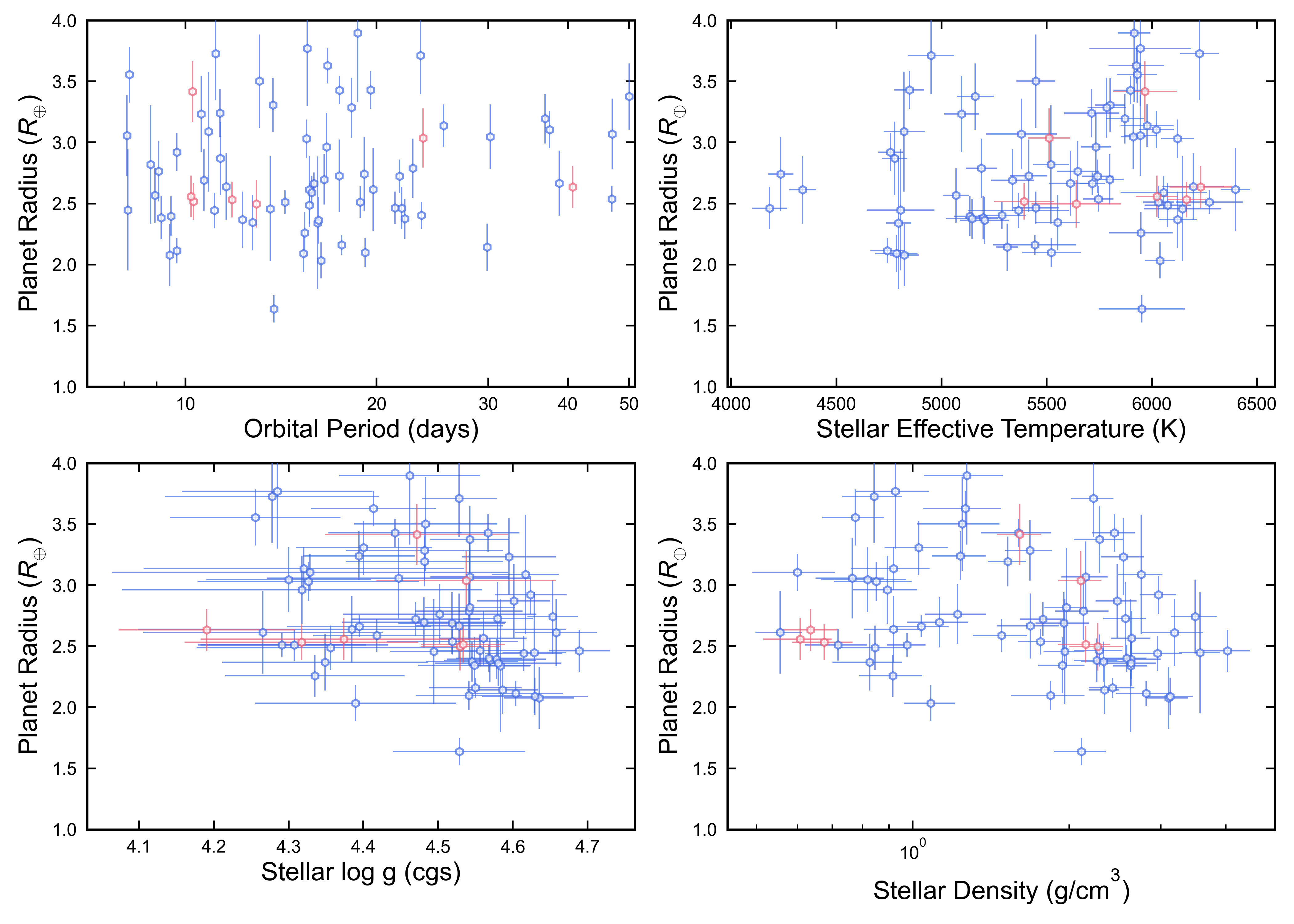}
    \caption{Host--planet properties for the $R_p<4\,R_\oplus$ subsample. Planet radius is shown as a function of orbital period (\textbf{upper left}), stellar effective temperature (\textbf{upper right}), stellar surface gravity $\log g$ (\textbf{lower left}), and stellar density $\rho_\star$ (\textbf{lower right}). Blue points denote the full small-planet sample, while red points highlight systems whose individual photoeccentric posteriors have eccentricity modes $e_{\rm mode}>0.5$. Error bars show the 68\% credible intervals.}
    \label{app:smallplanet}
\end{figure*}

\begin{figure*}
    \includegraphics[width=0.99\textwidth]{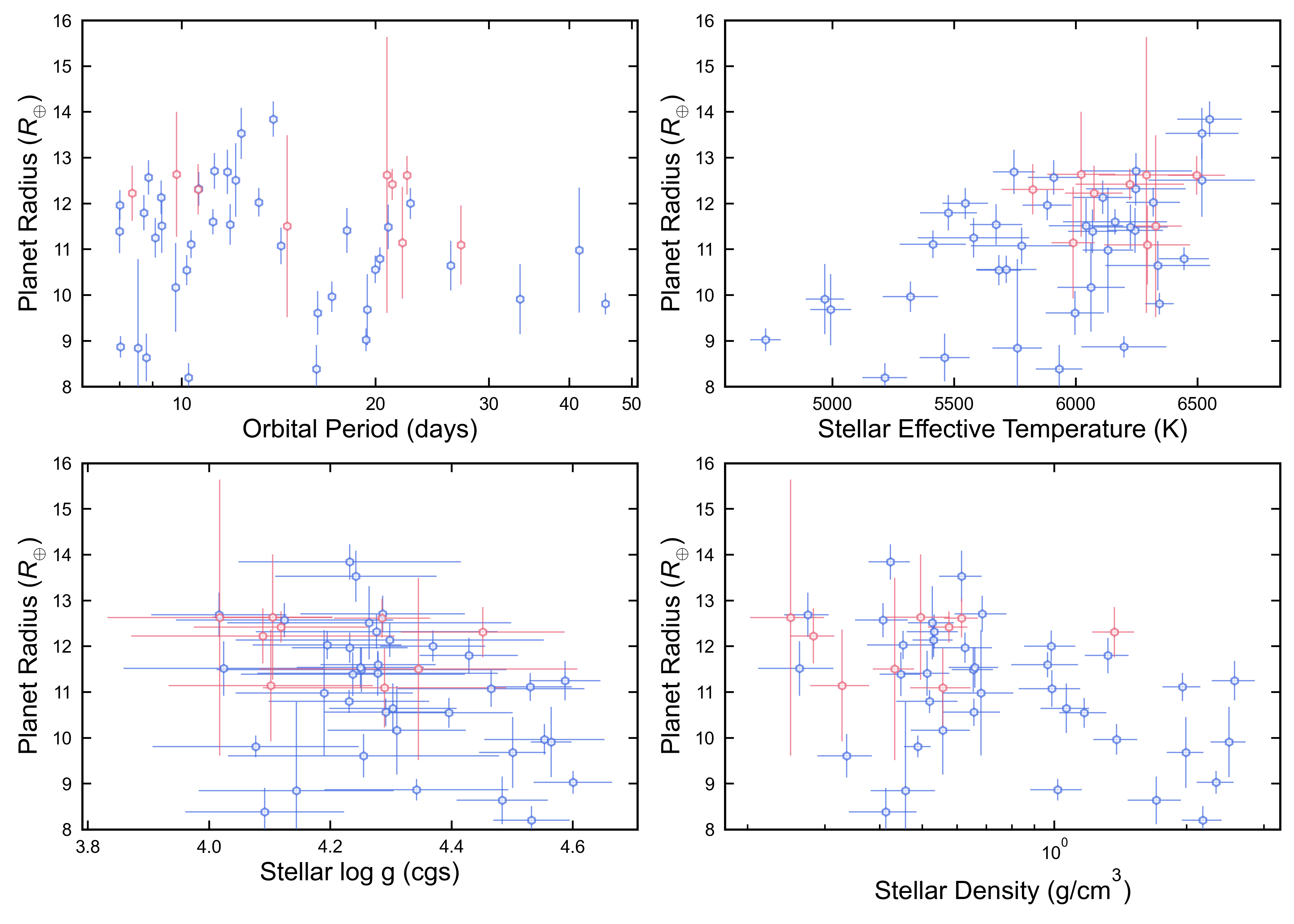}
    \caption{Same as Appendix \ref{app:smallplanet} but for giant planets ($R_p>8\,R_\oplus)$. }
    \label{app:giantplanet}
\end{figure*}

\begin{figure*}
    \includegraphics[width=0.99\textwidth]{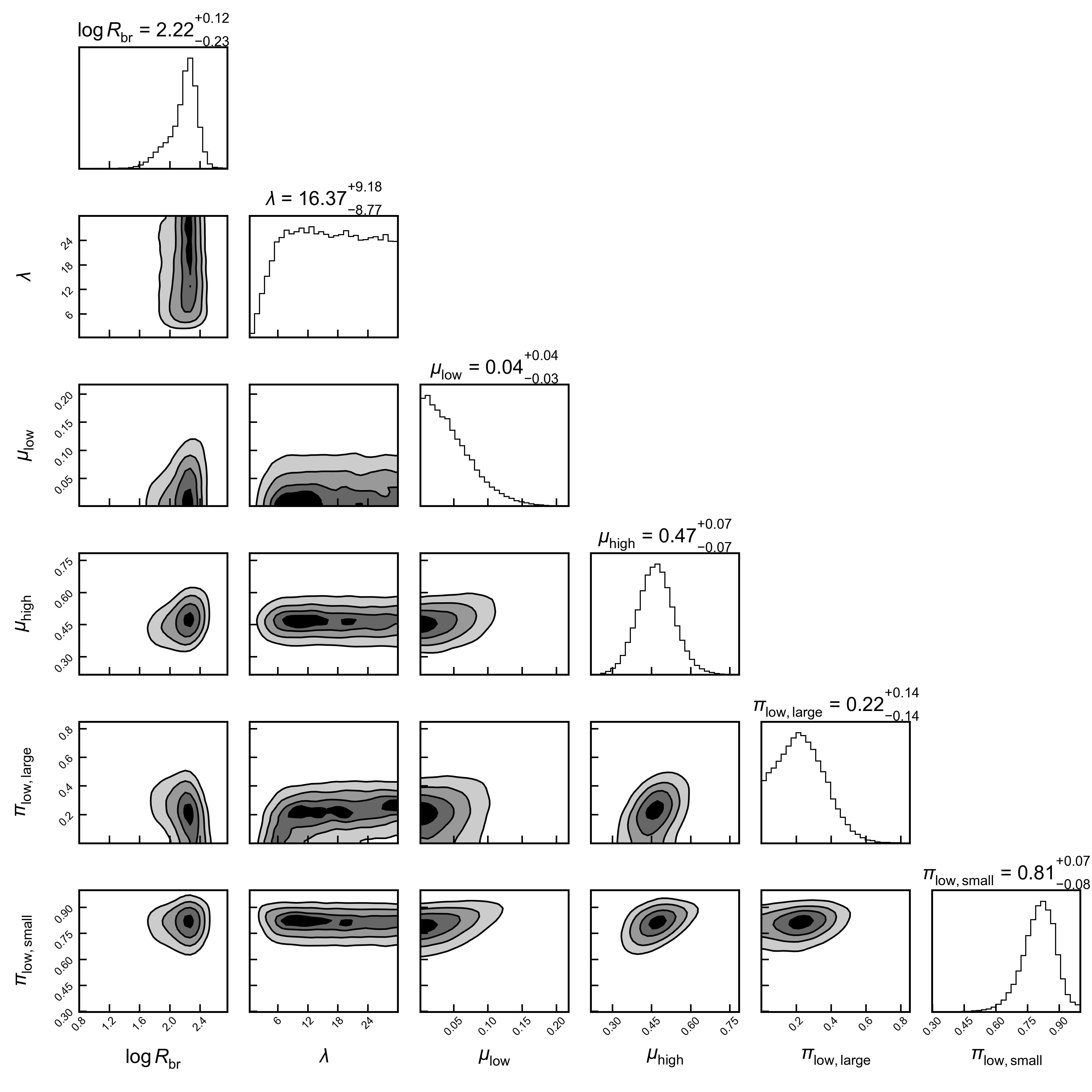}
    \caption{Corner plot of fiducial 3-stage sigmoid model.}
    \label{app:app_fig_corner_3stage}
\end{figure*}

\begin{figure*}
    \includegraphics[width=0.99\textwidth]{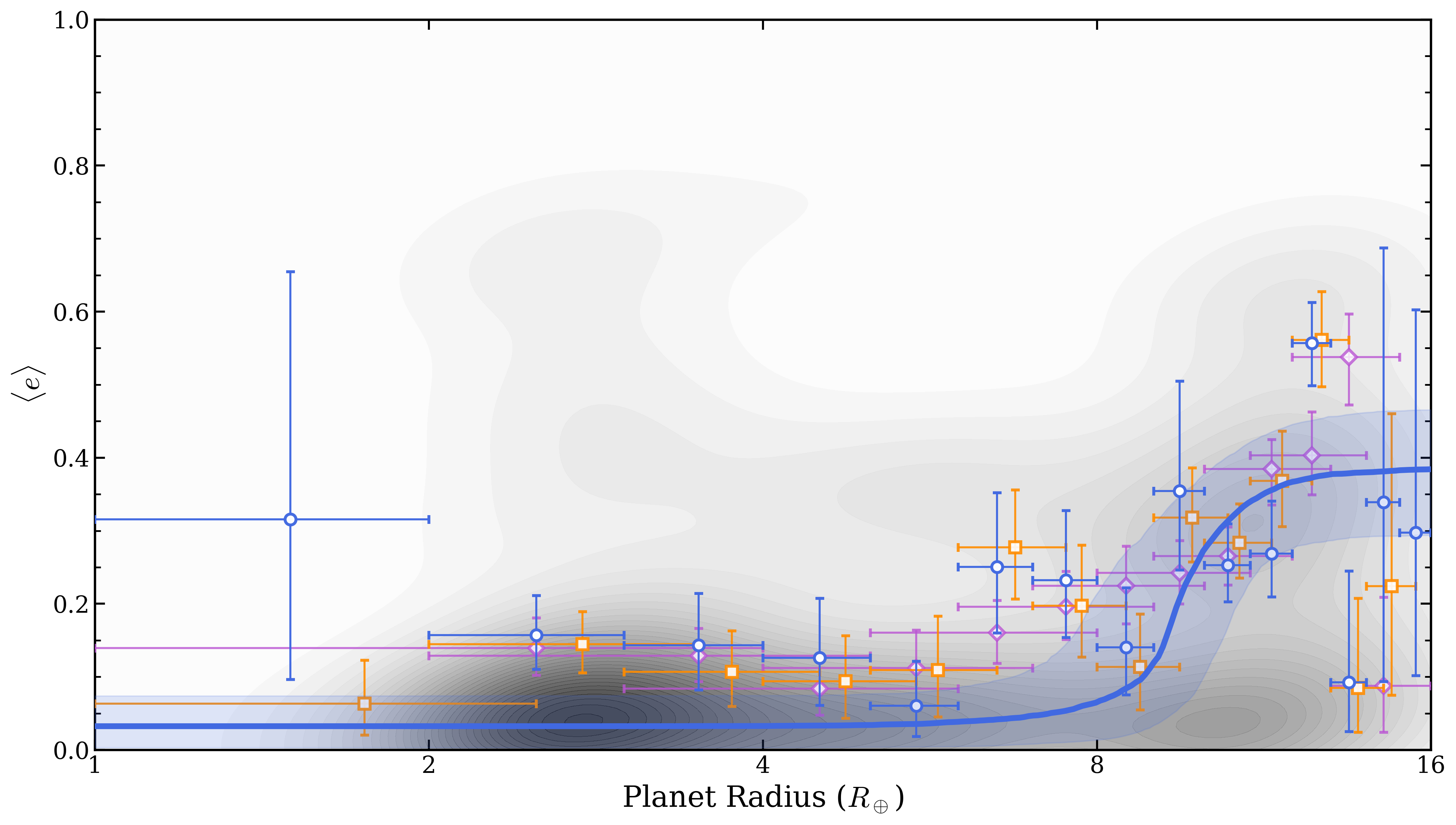}
    \caption{Same as Figure~\ref{fig:eccentricity_vs_radius} but including more discrete bin-size analyses: $\Delta 1.0\,R_p$ (blue), $\Delta 1.5\,R_p$ (orange), and $\Delta 3.0\,R_p$ (purple).}
    \label{app:bin_sizes}
\end{figure*}

\begin{figure}[h]
    \centering
    \includegraphics[width=0.8\textwidth]{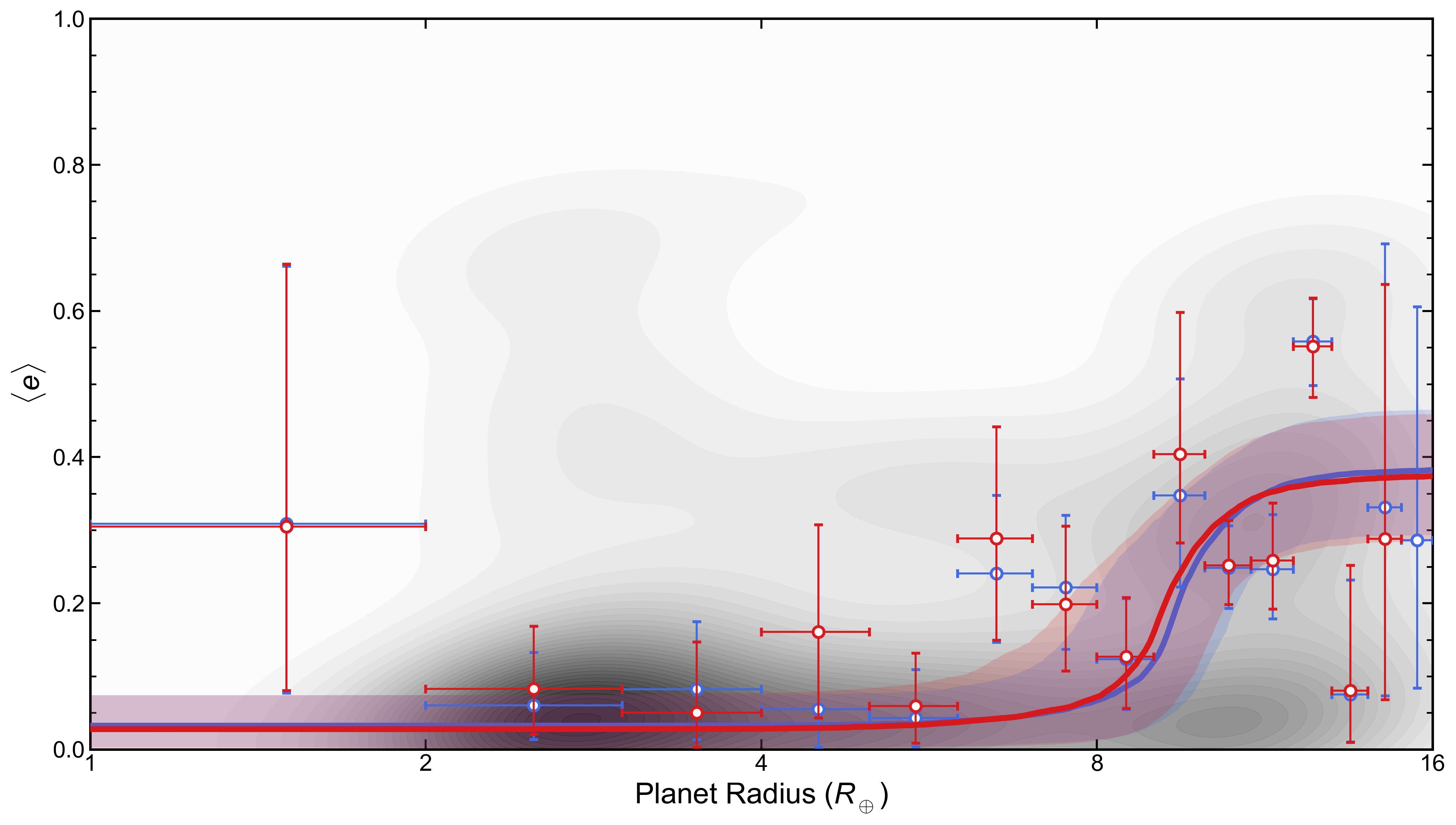}
    \caption{Same as Figure~\ref{fig:eccentricity_vs_radius}, except we also show the restricted confirmed and SG1 cleared planets in red. Open circles indicate single beta fits for discrete bin sizes, while the continuous curve is from the radius continuous model. One bin is missing between 15 and 16 Earth radii due to all planets being rejected in this cut. }
    \label{app:eccentricity_vs_radius_confirmed}
\end{figure}

\begin{figure*}
    \includegraphics[width=0.99\textwidth]{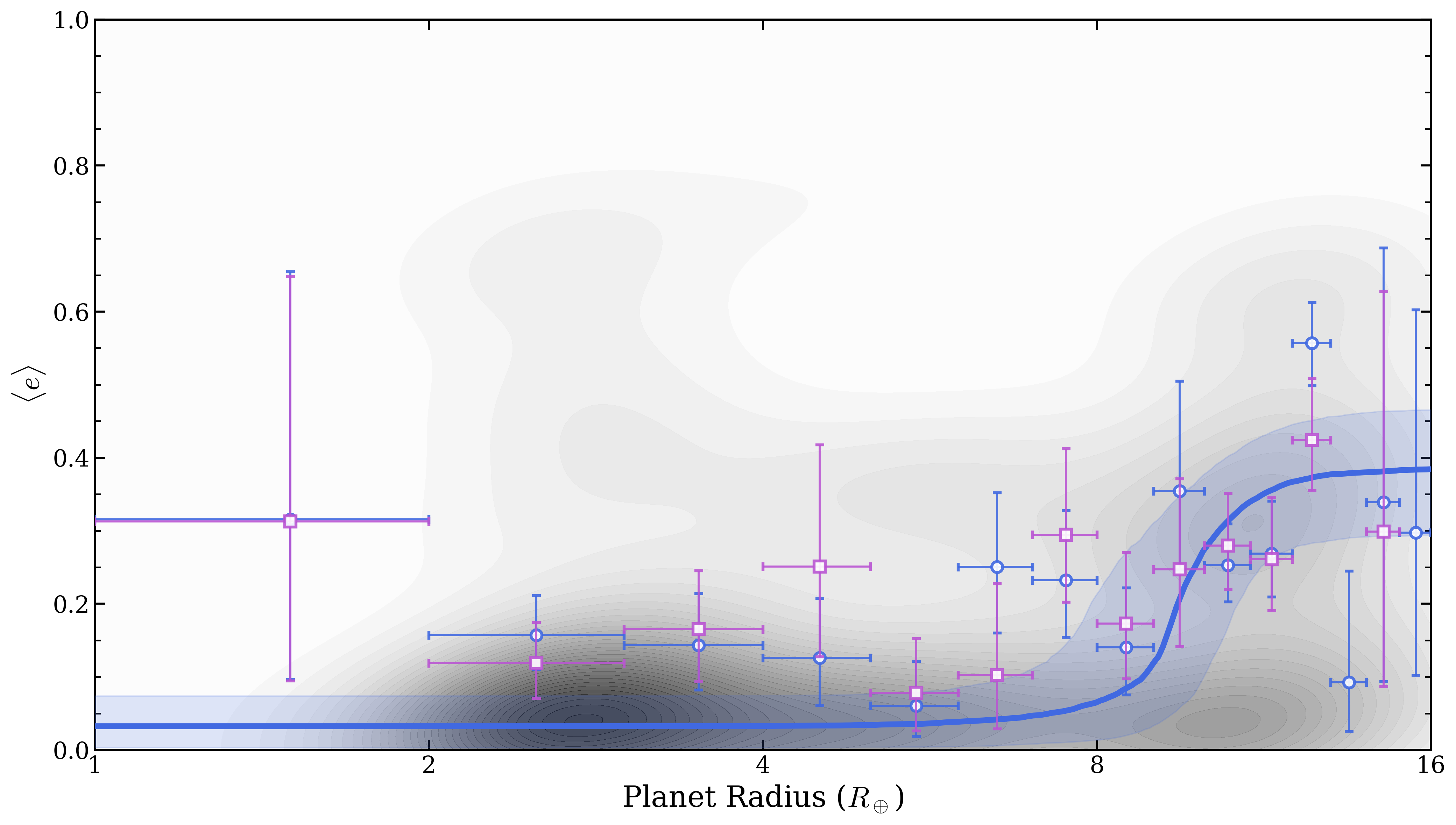}
    \caption{Same as Figure~\ref{fig:eccentricity_vs_radius} but showing the full sample (blue) compared to the GKM-type host targets only (purple), defined here as $T_{\mathrm{eff}}<6000$~K. Two regions of the GKM-type region are missing at 13-14$R_\oplus$ and 15-16$R_\oplus$ due to all planets in that region being around F-stars.}
    \label{app:stellar_dependence}
\end{figure*}

\bibliography{sample701}{}
\bibliographystyle{aasjournalv7}

\end{document}